# A Strange Detail Concerning the Variational Principle of General Relativity Theory

*Peter Ostermann\**

A mathematical complication due to an unnecessary formal assumption concerning the variational principle of general relativity theory, which apparently bothered Einstein and Hilbert, is shown and cleared up. Some historical confusion seems caused by the impossibility to use the conventional Euler-Lagrange formalism directly there, which even otherwise is nothing but one of various possible procedures to apply the superior principle of least action. Correspondingly to the absence of any direct calculation in the literature so far, only a numerical modification in parts – explicitly taken into account now after once mentioned by Hilbert without implementation – would allow to compute the fundamental Einstein tensor density from these authors' initial formulae, which must not be taken literally. Nevertheless adhering to a merely symbolic Euler-Lagrange formalism, this needs a clear distinction between 'component differentiation' and 'tensor differentiation' defined here. Various corresponding solutions are shown including the probably most natural one. Two of them are additionally verified in the detailed supplementary material appended to the electronic edition of the note.

### a) Introduction

There is a strange mathematical detail hidden in the variational principle of general relativity theory (GRT) as used by Einstein [1], Hilbert [2], Pauli [3], up to in our days e.g. Landau & Lifschitz [4], who claim Einstein's tensor density

$$\mathbf{E}_{ik} \equiv \sqrt{\mathbf{g}} E_{ik} \equiv \mathbf{R}_{ik} - \tfrac{1}{2}\mathbf{R} g_{ik} \qquad (1)$$

(with $R_{ik}$ that of Ricci and $R$ its trace) to result from Euler-Lagrange equations. The corresponding equations, however, – only of symbolic meaning in this case – must not be straightforwardly applied to the components of Einstein's symmetric covariant fundamental tensor $g_{ik}$ or of its contravariant counterpart $g^{ik}$, and their derivatives. These equations have either to be modified, or to be understood in a sense of artificial tensor operations defined here (numerically different from single-component operations of previously same notation, though).

### b) Concepts and notation

Given $g^{ik}$ as the contravariant fundamental tensor ($a, b, .. i, k, .. = 0..3$), its determinant is $|g^{ik}|$ with $\mathbf{g} \equiv -1/|g^{ik}|$ where by definition (using $\partial_l \equiv \partial/\partial x^l$ for partial differentiations with respect to the coordinates $x^l$) the expressions $g_l^{ik} \equiv \partial_l g^{ik}$ and correspondingly $g_{lik} \equiv \partial_l g_{ik}$ are mere abbreviations. Temporarily using any general Lagrangian density $\mathbf{L} \equiv \sqrt{\mathbf{g}}\, L$ of GRT and applying the superior principle of least action it has to be built

$$\delta \int \mathbf{L}\, dx = \int \delta \mathbf{L}\, dx = \int \left( L\, \delta\sqrt{\mathbf{g}} + \sqrt{\mathbf{g}}\, \delta L \right) dx, \qquad (2)$$

where $dx \equiv dx^0 dx^1 dx^2 dx^3$ and any bold faced symbols represent *densities*, which include always a first factor $\sqrt{\mathbf{g}}$ (except for $\mathbf{g}$ itself). Evaluating (2) it will be used the well-known relation

$$\delta \sqrt{\mathbf{g}} = -\tfrac{1}{2} \sqrt{\mathbf{g}}\, g_{ik}\, \delta g^{ik}. \qquad (3)$$

Already here, in contrast to the conventional handling of the Euler-Lagrange formalism it has to be realized the characteristic inequality

$$\delta \sqrt{\mathbf{g}} \neq (\partial \sqrt{\mathbf{g}} / \partial g^{ik})\, \delta g^{ik}$$

[cf. (24) below] as long as partial derivatives like $\partial/\partial g_{ik}$, $\partial/\partial g_{lik}$, $\partial/\partial g_{lmik}$, $\partial/\partial g^{ik}$, $\partial/\partial g_l^{ik}$, $\partial/\partial g_{lm}^{ik}$ are straightforwardly applied to corresponding tensor components. Throughout, 'straightforward' means that in particular Einstein's summation rule has to be used directly without occasional exceptions where appropriate. The exemplary inequality above is easily verified by mathematical software [at first a symmetric $g_{ik}$ may be defined including general components, then $g^{ik}$ is its inverse matrix (or the other way round)].

Though one might restrict the further treatment in principle to e.g. the contravariant components of the fundamental tensor $g^{ik}$ and its derivatives $g_l^{ik}$, $g_{lm}^{ik}$ only – where any appearing components $g_{(lm)ik}$ would have to be taken as single functions of the inverse tensor $g_{ab}$ of $g^{ik}$ [and correspondingly of $g_{(lm)}^{ik}$] – it will prove more convenient to allow even a mixed treatment dealing with $g^{ik}$ and $g_{lik}$, for example, or with $g^{ik}$, $g_{ab}$ and $g_{(lm)}^{ik}$ instead, where the tensors $g^{ik}$ and $g_{ab}$ may be varied independently at first, what would mean temporarily $\partial g^{ik}/\partial g_{ab} = \partial g_{ab}/\partial g^{ik} = 0$, before any interim differential $\delta g_{ab}$ may be converted to $\delta g^{ik}$ or vice versa in the end. Here in general it is

$$\delta[..] = \delta_\mathrm{I}[..] + \delta_\mathrm{II}[..] + \delta_\mathrm{III}[..] + \delta_\mathrm{IV}[..], \qquad (4)$$

where as usual the Lagrangian $L$ is assumed not to depend on higher than second derivatives of the fundamental tensor in this context. The necessary and sufficient presupposition concerning relation (4) is that $\delta[..]$ has to be a total differential with respect to the fundamental tensor (in both forms) and its partial derivatives, before making use of the well-known relation

$$\delta g_{ab} = -g_{ia} g_{kb}\, \delta g^{ik} \qquad (5)$$

\* Independent Research, Valpichlerstr. 150, 80689 Munich, GER
  Electronic address: peos@independent-research.org

resulting from $\delta(\delta_i^k) = 0 = \delta(g_{im} g^{mk})$. Correspondingly it is

$$g_{lik} = - g_{ia} g_{kb} g_l^{ab} \qquad (6)$$

resulting from $\partial_l \delta_i^k = 0 = \partial_l (g_{im} g^{mk})$. The required variations with respect to the covariant or contravariant fundamental tensor and their partial derivatives are involved according to

$$\delta_{\mathrm{I}}[..] = ... + \frac{1}{X} \frac{\partial [..]}{\partial g_{(l)}^{ik}} \delta g_{(l)}^{ik} + ..., \qquad (7)$$

$$\delta_{\mathrm{II}}[..] = ... + \frac{1}{X} \frac{\partial [..]}{\partial g_{(l)ik}} \delta g_{(l)ik} + ..., \qquad (8)$$

$$\delta_{\mathrm{III}}[..] = ... + \frac{1}{Y} \frac{\partial [..]}{\partial g_{(lm)}^{ik}} \delta g_{(lm)}^{ik} + ..., \qquad (9)$$

$$\delta_{\mathrm{IV}}[..] = ... + \frac{1}{Y} \frac{\partial [..]}{\partial g_{(lm)ik}} \delta g_{(lm)ik} + ..., \qquad (10)$$

with (*l*) a first facultative index of differentiation – where e.g. $\delta_{\mathrm{I}}[..] \equiv \delta_1[..] + \delta_2[..]$ may occur twice, with and/or without (*l*)-differentiation – while the pair of optional indices (*lm*) may represent second derivatives in addition. According to a statement by Hilbert [2] – with no explicit numerical implementation into corresponding relations there – two necessary assignments

$$X(ik) \equiv (1 + \mathrm{sign}|i - k|), \qquad (11)$$

$$Y(lm, ik) \equiv X(lm) \cdot X(ik), \qquad (12)$$

will be used in the following where $1/X$, $1/Y$ may be called 'Hilbert factors' existing due to the (*ik*)-symmetry of Einstein's fundamental tensor and the (*lm*)-symmetry of two-fold differentiations [as already above, the arguments of $X$, $Y$ as e.g. (*ik*), (*lm*) will be omitted where clear]. In contrast to a well-defined 'component differentiation' as usable straightforwardly, such Hilbert factors suggest the definition of a merely symbolic kind of 'tensor differentiation' always indicated by angle brackets $<..>$ as for example in

$$\left\langle \frac{\partial [..]}{\partial g_{(l)}^{ik}} \right\rangle \equiv \frac{1}{X} \frac{\partial [..]}{\partial g_{(l)}^{ik}}, \qquad (13)$$

$$\left\langle \frac{\partial [..]}{\partial g_{(l)ik}} \right\rangle \equiv \frac{1}{X} \frac{\partial [..]}{\partial g_{(l)ik}}, \qquad (14)$$

where only each right hand side applies to the components of corresponding expressions directly, while the 'artificial' *tensor constructs* $<..>$ respectively on the left will serve for calculations of true differentials $\delta(..)$ in the following. Obviously this allows to get quasi-familiar calculation rules for variations as in particular

$$\delta_{\mathrm{I}}[..] = ... + \left\langle \frac{\partial [..]}{\partial g_{(l)}^{ik}} \right\rangle \delta g_{(l)}^{ik} + ... \qquad (15)$$

$$\delta_{\mathrm{II}}[..] = ... + \left\langle \frac{\partial [..]}{\partial g_{(l)ik}} \right\rangle \delta g_{(l)ik} + ... \qquad (16)$$

which exemplary relations are equivalent by definition to (7), (8) above, now analogous to those for ordinary functions like e.g. $\delta f(x^i) = [\partial f(x^i)/\partial x^k] \delta x^k$.[1] – Special cases are

$$\delta g^{ab} = \Delta_{ik}^{ab} \delta g^{ik}, \qquad (17)$$

$$\left\langle \frac{\partial g^{ab}}{\partial g^{ik}} \right\rangle = \Delta_{ik}^{ab} = \frac{1}{X} \frac{\partial g^{ab}}{\partial g^{ik}}, \qquad (18)$$

$$\Delta_{ik}^{ab} \equiv \tfrac{1}{2} \left( \delta_i^a \delta_k^b + \delta_k^a \delta_i^b \right), \qquad (19)$$

or

$$\delta g_{cab} = \Delta_{cab}^{lik} \delta g_{lik} \qquad (20)$$

$$\left\langle \frac{\partial g_{cab}}{\partial g_{lik}} \right\rangle = \Delta_{cab}^{lik} \equiv \frac{1}{X} \frac{\partial g_{cab}}{\partial g_{lik}}, \qquad (21)$$

$$\Delta_{cab}^{lik} \equiv \tfrac{1}{2} \delta_c^l \left( \delta_a^i \delta_b^k + \delta_a^k \delta_b^i \right). \qquad (22)$$

Furthermore, as usual, it is presupposed the possible exchange of differentiation with variation

$$\delta(\partial_l ..) = \partial_l(\delta ..)$$

to reverse order. To demonstrate in particular the appropriate $<..>$ normalization as a convenient mathematical tensor artefact, this may be shown by verification of

$$\left\langle \frac{\partial g_{ab}}{\partial g^{ik}} \right\rangle = - \tfrac{1}{2} \left( g_{ia} g_{kb} + g_{ka} g_{ib} \right) \qquad (23)$$

using relation (13) where in this case the facultative index (*l*) has to be left out and a direct calculation of $1/X \cdot (\partial g_{ab}/\partial g^{ik})$ yields the result $-\tfrac{1}{2}(g_{ia} g_{kb} + g_{ka} g_{ib})$ on the right.

---

[1] In a footnote of [4], Landau & Lifschitz emphasize that the notation of differentiations "$\partial F/\partial g_{ik}$ (*F* a function of $g_{ik}$)" with respect to the "components of the symmetric tensor $g_{ik}$" be only of symbolic character in a sense (thereby correctly addressing the feature described by the 'Hilbert factors' introduced above), while here in contrast (to the symbolic $<(..)>$ operations) just the 'component differentiations' (..) are explicitly verified to apply straightforwardly. In another context, the general importance of "a clear distinction between equations that hold between tensors, and equations for their components ..." is emphasized by Wald [5].

Now, without a circumstantial straightforward calculation, it is allowed to conclude (23) as suggested by the valid relation $\delta g_{ab} = -g_{ia}g_{kb}\,\delta g^{ik}$ of (5), where obviously [2] the non-proper tensor quantity '$\delta g_{ab}/\delta g^{ik}$' should be equivalent to the tensor construct $\langle\partial g_{ab}/\partial g^{ik}\rangle$ above. Correspondingly, as another example for such an apparent equivalence

$$'\frac{\delta[..]}{\delta g^{ik}}' \to \left\langle\frac{\partial[..]}{\partial g^{ik}}\right\rangle$$

now from (3) it is also directly suggested the valid relation

$$\left\langle\frac{\partial\sqrt{\mathbf{g}}}{\partial g^{ik}}\right\rangle = -\tfrac{1}{2}\sqrt{\mathbf{g}}\,g_{ik} \qquad (24)$$

though any division by matrix quantities (as e.g. by $\delta g^{ik}$ above) would not at all be a legitimate tensor operation if understood literally [this relation (24) has been explicitly verified by direct calculation with respect to general symmetric $g_{ik}$-components again according to the definition (13), where necessarily (11) has to be taken into account]. A comparison proves the unnumbered inequality below (3) definitely.

The standard Euler-Lagrange formalism would presuppose the equality of expressions like

$$\delta f(x^i) = \frac{\partial f(x^i)}{\partial x^k}\delta x^k, \qquad (25)$$

which is not given if $f(x^i)$ was replaced by the fundamental tensor $g_{ik}$ or its inverse $g^{ik}$. Evidently, from direct calculation it turns out that for example

$$\delta g_{ab} \neq \frac{\partial g_{ab}}{\partial g^{ik}}\delta g^{ik},$$

since according to (7), (13), (23), here the right hand side equals $X$ times the right hand side of (5). Dealing with tensors it makes a difference whether this is done with regard to their single components or with regard to whole tensors only. If applied as usual, exactly the presupposition above seems not fulfilled. Nevertheless, if trying to adhere to a symbolic Euler-Lagrange formalism, then it has proved appropriate to define e.g.

$$\delta_1[..] = \left\langle\frac{\partial[..]}{\partial g^{ik}}\right\rangle\delta g^{ik} \equiv \frac{1}{X}\frac{\partial[..]}{\partial g^{ik}}\delta g^{ik}, \qquad (26)$$

---

[2] The explicitly ($ik$)-symmetrized form of $g_{ia}g_{kb}$ as used in (23) is unnecessary in relation (5) because of the factor $\delta g^{ik}$ there. Relation (23) has been shown in Section III of v3 (with a different numerical assignment of $X$ and of what has already been called 'Hilbert factor' there).

---

according to (15), together with corresponding relations according to (16) - (18) above. Doing so, it is clear that it may be only a question of convention which of both kinds of derivatives with respect to $g_{ik}$-quantities in (26) is set into angle brackets. In any case, however, it is necessary to avoid potential confusion by a clear distinction of various symbols and operations.

Taken together, the conventional Euler-Lagrange formalism is nothing but in many cases the simplest procedure *among others* to apply the principle of least action. Exactly the latter superior principle definitely includes the straightforward viable alternatives of the next two sections where, if appropriate, the variations may also be done with respect to both the covariant and the contravariant representations of Einstein's metric fundamental tensor independently at first. In general, any total differential $\delta Z$ will come out with same result, no matter which portions $A, B \ldots N$ may be taken internally together, if only these portions build up the complete expression $Z(A, B \ldots N)$ in question, which then may be varied with respect to $A, B \ldots N$, before any relations between the differentials $\delta A, \delta B \ldots \delta N$ have to be evaluated after all. In case of this note, $A, B, C$ may represent $\sqrt{\mathbf{g}}$, $g^{ik}$, $g_{l\,ik}$, and $Z$ the Lagrangian $\mathbf{L}$ for example.

'Some additional relations and calculation rules' are given in Section I of the supplementary material appended to the electronic edition of this note.

### c) Einstein's treatment

Concluding his treatment of GRT's variational principle in "*Hamiltonsches Prinzip und allgemeine Relativitätstheorie*" [1] of 1916, Einstein claimed the tensor density (1) to result from the formula

$$\frac{\partial\mathbf{G}}{\partial g^{ik}} - \partial_l\left(\frac{\partial\mathbf{G}}{\partial g^{ik}_l}\right). \qquad (27)$$

As is well-known, the Lagrangian $\mathbf{G}$ of the gravitational field may be written in the form

$$\mathbf{G} \equiv \sqrt{\mathbf{g}}\,G, \qquad (28)$$

where

$$G \equiv \left(g^{um}g^{sv}g^{rw}\right)\{\Gamma_{v,ur}\Gamma_{w,ms} - \Gamma_{v,um}\Gamma_{w,sr}\}, \qquad (29)$$

and Christoffel's symbols of first kind are

$$\Gamma_{l,ik} \equiv \tfrac{1}{2}(g_{kli} + g_{ilk} - g_{lik}). \qquad (30)$$

That something might be problematic with expression (27) above, is indicated by the peculiarity, that there actually seems not to exist any straightforward calculation of Einstein's tensor density $\mathbf{E}_{ik}$ (1) from this ansatz. It will be shown here that (27) – if assumed to equal (1) – suffers at least from a misleading mathematical notation.

On the other hand, remarkably enough, this obstacle is overcome by authors from Weyl [6] to e.g. Weinberg [7] or Wald [5], who treat the variational principle of GRT alternatively in that they calculate $\mathbf{E}_{ik}$ without using partial derivatives like in particular $\partial/\partial g^{ik}$, $\partial/\partial g_l{}^{ik}$ exclusively (or at all). Starting from the proof that $\delta \int \mathbf{G} dx$ is a 4-dimensional scalar, Weyl directly converted

$$\delta \int \mathbf{G} dx = \int \delta \mathbf{G} dx = \int \mathbf{E}_{ik} \delta g^{ik} dx \qquad (31)$$

in a most elegant way. – One reason why in contrast to (31) relation (27) does not work straightforwardly, is that $\mathbf{G}$ cannot be written without $g_{ik}$ as a complete self-contained algebraic function of pure 'tensor quantities' like $g^{ik}$ and its direct partial derivative $g_l{}^{ik}$ only. This, however, has been effectively presupposed by Einstein and others according to the usual physical treatment of variational principles so far, though such a presupposition would not literally apply here.[3] Instead of that unnecessary requirement, $G$ does not exist in a corresponding form other than of single components only.

The relevant criterion, however, is to find the total differential yielding $\mathbf{E}_{ik}$ according to (31). To this end, the chain rule of differentiation has to be consequently applied in any variational principle, which in special cases is implying the well-known Euler-Lagrange formalism among other procedures, for example.

Here, from relations (28), (29), (30) it is clearly seen that it is $\mathbf{G} \equiv \mathbf{G}(\sqrt{\mathbf{g}}, g^{ik}, g_{lik})$ with $G$ algebraically composed of the *contravariant* tensor $g^{ik}$ and the first derivatives $g_{lik}$ of the *covariant* tensor $g_{ik}$ directly. Thus – according to the definition of symbolic 'tensor differentiation' indicated by angle brackets <..> whose actual calculation rules are given in the previous section – at first it is

$$\delta \mathbf{G} = \left\langle \frac{\partial \mathbf{G}}{\partial g^{ik}} \right\rangle \delta g^{ik} + \left\langle \frac{\partial \mathbf{G}}{\partial g_{lab}} \right\rangle \delta g_{lab}. \qquad (32)$$

Inserting this simplest starting relation into (31), then – after partial integration with all variations presupposed to vanish at the integrals' boundaries as well as making use of (5) – now it follows by comparison with the right hand side of (31) above:

$$\mathbf{E}_{ik} = \left\langle \frac{\partial \mathbf{G}}{\partial g^{ik}} \right\rangle + g_{ia} g_{kb} \partial_l \left\langle \frac{\partial \mathbf{G}}{\partial g_{lab}} \right\rangle. \qquad (33)$$

One may directly verify – though a somewhat laborious procedure (s. supplementary material) – that in contrast to (27) this relation immediately works, calculating Einstein's tensor density from relation (33) straightforwardly.

Instead of Einstein's $\mathbf{t}_\sigma{}^\nu$ in [1]/(20) the energy-momentum pseudo-tensor of the gravitational field, if derived correspondingly to (33), equals

$$\mathbf{e}_i^k \equiv \frac{1}{2} \left\{ \mathbf{G} \delta_i^k - g_{ilm} \left\langle \frac{\partial \mathbf{G}}{\partial g_{klm}} \right\rangle \right\}, \qquad (34)$$

except for a constant factor (in accordance to its traditional expression after all). Here in particular relation (24) has been used as will be done in the following.

Furthermore, relation (33) is not the only valid representation according to (31). If one insisted on handling $\mathbf{G}$ as a function $\mathbf{G}[g^{ik}, g_{lik} := -g_{ia} g_{kb} g_l{}^{ab}$ according to (6)], one may also start with

$$\delta \mathbf{G} = \left\langle \frac{\partial \mathbf{G}^{(ik)}}{\partial g^{ik}} \right\rangle \delta g^{ik} + \left\langle \frac{\partial \mathbf{G}}{\partial g_l{}^{ik}} \right\rangle \delta g_l{}^{ik} + \left\langle \frac{\partial \mathbf{G}_{(ab)}}{\partial g_{ab}} \right\rangle \delta g_{ab} \qquad (35)$$

instead of (32) at first, and find

$$\mathbf{E}_{ik} = \left\langle \frac{\partial \mathbf{G}^{(ik)}}{\partial g^{ik}} \right\rangle - \partial_l \left\langle \frac{\partial \mathbf{G}}{\partial g_l{}^{ik}} \right\rangle - g_{ia} g_{kb} \left\langle \frac{\partial \mathbf{G}_{(ab)}}{\partial g_{ab}} \right\rangle \qquad (36)$$

where $\partial \mathbf{G}^{(ik)} / \partial g^{ik}$ and $\partial \mathbf{G}_{(ab)} / \partial g_{ab}$ obviously mean independent partial derivations of the same Lagrangian $\mathbf{G}$ with respect to the contravariant and covariant tensor factors appearing there now [relation (5) has been used again]. It is unnecessary to discuss (36) as compared to (27) here at length since – in contrast to the historical development – there is the most natural expression (33) leading to the Einstein tensor density by straightforward direct calculation.[4]

In 1915, discussing expressions analogous to (27) and (31) – s. Doc.s 60, 62, 64, 66, 67, 69, 71, 74, 75, 77, 78, 80 of [9] CPAE Vol. **8**A – Levi-Civita irrefutably proved as a 'general theorem' in his equation (30) of Doc. 67 that expression (27) above would not be a tensor density. Though, the historical peculiarity is that he used Einstein's earlier Lagrangian

$$\sqrt{\mathbf{g}} H = -\sqrt{\mathbf{g}} g^{um} \tilde{\Gamma}_{ru}^s \tilde{\Gamma}_{sm}^r \qquad (37)$$

taken from relation (78) in [10] according to the corresponding footnote 1 in that article (here respectively distinguished by the usage of tildes) where, however, the quantities $\tilde{\Gamma}_{kl}^i \equiv 1/2 \, g^{im} g_{lmk} \neq \Gamma_{kl}^i \equiv g^{im} \Gamma_{m,kl}$ are different from the Christoffel symbols of second kind. Otherwise expression (37) would have been a restriction of the full **G** with sign changed given in (28), (29), where the missing summand would equal zero in case of any coordinate choice implying $\sqrt{g} = 1$. Though the mistaken ansatz (37) has been the actual reason in this context, even otherwise expression (27) – if taken literally – would have hardly yielded the Einstein tensor because of its inappropriate notation addressed above.

---

[3] That such a presupposition may be misleading was found applying the variational principle for different approaches, which led in particular to [8] (one $\sqrt{\mathbf{g}}$ typo in the Klein-Gordon equation deduced there).

[4] Only using the angle-bracket normalization of the previous section including the particular tensor construct $<\partial g_{ab}/\partial g^{ik}>$ to equal $-g_{ia} g_{kb}$ according to (23) as only suggested by the valid relation (5), then relation (36) above could be formally reduced to Einstein's expression (27), as shown in the supplementary material [s. in particular Section IV "Exposing the problematic form …"].

Levi-Civita's disproof of Einstein's claim has much later been pointed out by Cattani & De Maria in [11], who found that "After many fruitless attempts (…) Einstein was obliged for the first time to admit that both the proof of that theorem and its consequences were not correct". Furthermore Pais [12], Norton [13], Howard & Stachel [14], Renn et al. 2007 [15], Renn & Sauer [16] have reconstructed how after a "a chain of erroneous paths" Einstein came up with his final theory. At last, in particular Janssen & Renn 2007 [17] have obviously tracked and enlightened Einstein's historical way from the Zurich notebook to his final equations almost completely.[5] Already here, Einstein had effectively presupposed the unnecessary assumption that his $H$, corresponding to (37), was composed of $g^{ik}$, $g_l^{ik}$ only. This obstacle – besides the inappropriate **H** itself – still stayed present through the whole following correspondence, while the components $\tilde{\Gamma}^i_{kl} \equiv \tfrac{1}{2} g^{im} g_{lmk}$ obviously depend on $g^{ik}$, $g_{lik}$, where $g_{lik}$ refers to the original covariant fundamental tensor of GRT directly.

After Levi-Civita's striking disprove in Doc. 67, Einstein in Doc. 69 found himself temporarily at such a defensive position that he unacceptably argued *"My proof fails just in the special case you dealt with"* [6] before he concluded the same letter with *"In general, however, (…) my proof holds rightly …"* [7]. In an appendix he wrote: *"I have not yet experienced such an interesting correspondence"* [8].

The problems with expression (27) – inaccurate for $\mathbf{E}_{ik}$ if taken literally – do not at all affect the validity of Einstein's wonderful equations of course. That they, however, might have confused Einstein even while writing his fundamental 1916 paper [18] on GRT, can be seen from a manuscript published posthumously (s. Doc. 31 of [19] CPAE Vol. **6**), which *"was originally intended as §14 of ... and later as an appendix to ..."*, before it was left out in the end.[9] In that manuscript he tried to handle the variational principle without the restriction $\sqrt{g} = 1$ (otherwise mostly used), but – adhering to the problematic ansatz (27) above – he obviously did not succeed to his complete satisfaction.[10]

It took about seven months, before Einstein came back with "Hamilton's Principle and the General Theory of Relativity" [1], where he refers to the achievements of Hilbert [2] and Lorentz [20] explicitly[11], s. also Doc.s 183/84 of [9]. Thus, he retained expression (27), apparently without further trying to verify it by direct calculation which, with hindsight, would effectively have led to the unnecessarily complicated expression

$$\left\langle \frac{\partial \mathbf{G}^{(ik)}}{\partial g^{ik}} \right\rangle - \partial_l \left\langle \frac{\partial \mathbf{G}}{\partial g_l^{ik}} \right\rangle + \left\langle \frac{\partial g_{ab}}{\partial g^{ik}} \right\rangle \left\langle \frac{\partial \mathbf{G}_{(ab)}}{\partial g_{ab}} \right\rangle , \quad (38)$$

which with respect to (23) is equivalent to (36) after all. The ironic background, however, is that Hilbert in his famous "first note" [2] – using the Riemannian curvature scalar density **R** instead of **G** – had claimed another expression analogous to (27), which in this literal form has been inappropriate for a straightforward derivation of $\mathbf{E}_{ik}$, too, *"… readily without calculation …"* [12]. This note is dated of 1915 but published in 1916 with several changes, though.

---

[5] The 1915 Einstein-Levi-Civita correspondence would not have been addressed here again if there was not that additional mathematical complication above.

[6] *"Mein Beweis versagt gerade in dem von Ihnen behandelten Spezialfalle"*, [9] CPAE Vol. **8**A, Doc. 69 – (all German citations translated by author).

[7] *"Im allgemeinen wird sich aber ... Dann besteht mein Beweis zu Recht ..."*, [9] CPAE Vol. **8**A, Doc. 69.

[8] *"Eine so interessante Korrespondenz habe ich noch nicht erlebt"*, [9] CPAE Vol. **8**A, Doc. 69. This amicable controversy – now because of the problematic Euler-Lagrange approach, too – might appear in some additional light again.

[9] It may be another case for Einstein's unique intuition that he left out this '§ 14' appendix, though – explicitly using an unnecessary presupposition again – he apparently arrived with the desired result. The reason finally to omit this problematic derivation, might go back to Levi-Civita's disproof of the assumed tensor character of (27) – previously applied to (37) – which certainly had left a permanent impression though Einstein now used his final **G** (28), (29),(30) instead of the incomplete **H** (37) of 1914 before. The context of relations (80) to (83) there – and in particular the first six text lines of §3 of this omitted appendix, placed around two freestanding expressions – contains Einstein's conclusion of (27) above, and has been obviously transferred to the treatment in [1] later on.

[10] Though in an analogous form to (27) as a function of $g^{ik}$ and $g_l^{ik}$, later in [1] also the energy-momentum conservation law is derived as stated by the editors' note 6 of Doc. 31, CPAE Vol. **6**, who pointed out a letter to de Donder of 23 July 1916 where '… Einstein admitted a few months later he did not carry out the calculation'. Also remarkably, Einstein wrote there that the variation of the action integral may confidentially be done with respect to "$g_{\mu\nu}$ (or $g^{\mu\nu}$)", what – as can be seen from e.g. the Lagrangian of Maxwell's electromagnetic field – in general does not necessarily yield identical results. He wrote „*Man kann als Variationsprinzip ruhig ansetzen* $\delta\int(...)$, *wobei bezüglich der* $g_{\mu\nu}$ *(oder* $g^{\mu\nu}$*) zu variieren ist …*", two lines before he continued: „*Durchgeführt habe ich die etwas langwierige Berechnung der* $\mathbf{t}_\sigma^\nu$ = (…) *nicht.*" – For the energy-momentum conservation treatment including the de 1916 Donder episode s. Janssen & Renn 2007 again.

[11] "Recently H. A. Lorentz and D. Hilbert succeeded in …" [„*In letzter Zeit ist es H. A. Lorentz und D. Hilbert gelungen …*"], which introduction supports the conclusion that he had abandoned his own corresponding attempts. Regarding Doc. 41, CPAE Vol. 6, there in particular also footnote 2, p. 1113 is of importance {which by the way had obviously been overseen in the Editors' note [9] of Doc. 31}.

[12] „*… wie leicht ohne Rechnung aus der Tatsache folgt …*". Later on, in Pauli's book [3] it reads correspondingly: "*An explicit

**d) Hilbert's treatment**

A reason why Hilbert would have faced unnecessary complications up to probably an effective impracticability, too, is the unnecessary presupposition again that Riemann's curvature scalar should be handled as if algebraically composed of the pure tensor quantity $g^{ik}$ and its direct partial derivatives $g_l^{ik}$, $g_{lm}^{ik}$ completely. Although this difficulty is obvious from the previous section, now Hilbert's starting expression for the Einstein tensor density – which has not been worked out – may be confronted with a straightforwardly practicable treatment in detail.

As an extension of expression (27), he explicitly claimed the Einstein tensor density $\mathbf{E}_{ik}$ to result from his formula

$$\frac{\partial \mathbf{R}}{\partial g^{ik}} - \partial_l \left( \frac{\partial \mathbf{R}}{\partial g_l^{ik}} \right) + \partial_{lm} \left( \frac{\partial \mathbf{R}}{\partial g_{lm}^{ik}} \right), \quad (39)$$

where he used the full Riemannian curvature scalar density which may be written here

$$\mathbf{R} \equiv \sqrt{\mathbf{g}}\, g^{um} \{ \partial_v \Gamma^v_{um} - \partial_m \Gamma^v_{uv} \} - \mathbf{G} \quad (40)$$

(his $\mathbf{K} \equiv -\mathbf{R}$) in the variational principle instead of Einstein's $\mathbf{G}$ only. To show that analogously to the argumentation above, $\mathbf{R}$ is a complete function of $\sqrt{\mathbf{g}}$, $g^{ik}$, $g_{lik}$, $g_{mlik} \equiv \partial_m g_{lik}$, the term in curled brackets may be written in the form

$$\{ \partial_v \Gamma^v_{um} - \partial_m \Gamma^v_{uv} \} = g^{sv} (X_{umsv} + Y_{umsv}), \quad (41)$$

where

$$\begin{aligned} X_{umsv} &= (\partial_v \Gamma_{s,um} - \partial_m \Gamma_{s,uv}), \\ Y_{umsv} &= g^{rw} (g_{mvw} \Gamma_{s,ur} - g_{rvw} \Gamma_{s,um}). \end{aligned} \quad (42)$$

It is well-known, that $\delta \int \mathbf{R} dx$ is reducible by partial integration to $\delta \int \mathbf{G} dx$, presupposing that as usual all variations vanish at the integration limits (see e.g. [4]). Nevertheless, taking Hilbert's full action integral without this reduction to (31) exactly as it is, then in

$$\delta \int \mathbf{R} dx = \int \delta \mathbf{R} dx = \int \mathbf{E}_{ik} \delta g^{ik} dx \quad (43)$$

---

*evaluation now shows that ...*" {(27) should equal (1)}, where this author – besides a cross-reference to Palatini's [21] alternative procedure – points to Weyl's [6] treatment who, however, has done his actual calculation directly based on relation (31) instead of (27). Similarly Landau & Lifschitz [4] turn the table in that they calculate the Einstein tensor density $\mathbf{E}_{ik}$ from (43) like Weinberg did in [7], before they – however – only *conclude* without direct calculation that this should equal expression (27) again (s. also Footnote 1 above). The actual principle of least action has been also applied by e.g. Wald [5], Misner, Thorne, & Wheeler [22], and others without using Einstein's or Hilbert's relation (27), (39) explicitly.

with $\mathbf{R} \equiv \mathbf{R}(\sqrt{\mathbf{g}}, g^{ik}, g_{lik}, g_{mlik})$ given by (40), (41), (28), it is

$$\delta \mathbf{R} = \left\langle \frac{\partial \mathbf{R}}{\partial g^{ik}} \right\rangle \delta g^{ik} + \left\langle \frac{\partial \mathbf{R}}{\partial g_{lab}} \right\rangle \delta g_{lab} + \left\langle \frac{\partial \mathbf{R}}{\partial g_{lmab}} \right\rangle \delta g_{lmab}. \quad (44)$$

Thus the variational principle (43) results by partial integration in the working relation

$$\mathbf{E}_{ik} = \left\langle \frac{\partial \mathbf{R}}{\partial g^{ik}} \right\rangle + g_{ia} g_{kb} \partial_l \left[ \left\langle \frac{\partial \mathbf{R}}{\partial g_{lab}} \right\rangle - \partial_m \left\langle \frac{\partial \mathbf{R}}{\partial g_{lmab}} \right\rangle \right] \quad (45)$$

analogously to relation (33) above (verified by direct calculation, too). – It seems an interesting aspect in addition, that the use of $\mathbf{R}$ instead of $\mathbf{G}$ is extending the expression (34) by

$$\Delta \mathbf{e}_i^k \equiv \tfrac{1}{2} \{ (\mathbf{R} - \mathbf{G}) \delta_i^k + \partial_i [\sqrt{\mathbf{g}} g_l^{kl} + 2 g^{kl} \partial_l \sqrt{\mathbf{g}}] \}, \quad (46)$$

yielding another form $\mathbf{e}_{i\,(\mathbf{R})}^k \equiv \mathbf{e}_i^k + \Delta \mathbf{e}_i^k$ according to (51) below. The further relation $\partial_k (\Delta \mathbf{e}_i^k) \equiv 0$ applying here, means that there is no additional exchange of energy or momentum, though.[13]

And again, converting $\mathbf{R}$ of (40), (42) into a function $\mathbf{R}(g^{ik}, g_{lik} := -g_{ia} g_{kb} g_l^{ab}, g_{lmik} := ...)$, then instead of (45) one may also find $\mathbf{E}_{ik}$ equal to

$$\left\langle \frac{\partial \mathbf{R}^{(ik)}}{\partial g^{ik}} \right\rangle - \partial_l \left\langle \frac{\partial \mathbf{R}}{\partial g_l^{ik}} \right\rangle + \partial_{lm} \left\langle \frac{\partial \mathbf{R}}{\partial g_{lm}^{ik}} \right\rangle - g_{ia} g_{kb} \left\langle \frac{\partial \mathbf{R}_{(ab)}}{\partial g_{ab}} \right\rangle, \quad (47)$$

thus completing Hilbert's (39) explicitly. If, however, Hilbert's statement concerning ($ik$) and ($lm$)-symmetries had indicated that he understood his expression (39) in the sense of merely symbolic 'tensor differentiations' as defined above, then this should have practically led to

$$\left\langle \frac{\partial \mathbf{R}^{(ik)}}{\partial g^{ik}} \right\rangle - \partial_l \left\langle \frac{\partial \mathbf{R}}{\partial g_l^{ik}} \right\rangle + \partial_{lm} \left\langle \frac{\partial \mathbf{R}}{\partial g_{lm}^{ik}} \right\rangle + \left\langle \frac{\partial g_{ab}}{\partial g^{ik}} \right\rangle \left\langle \frac{\partial \mathbf{R}_{(ab)}}{\partial g_{ab}} \right\rangle \quad (48)$$

instead of (47). In this case, however, particularly relation (23) above would have been necessary to derive Einstein's tensor density (1), while – taken literally – a component differentiation '$\partial g_{ab}/\partial g^{ik}$' $\neq -g_{ia} g_{kb}$ would have disproved (39) in old notation.

Concerning (45) above, the derivation is done in more detail as follows. From (44) a repeated partial integration yields

---

[13] In this context, the same lack is obviously questioning any additional exchange of angular momentum in case of the well-known symmetric alternative pseudo-tensors of [4], [7], too, thus reducing their attraction significantly.

$$\delta \int \mathbf{R}\, dx = \int \{...\}\, dx + \int \partial_l \mathbf{Q}^l\, dx , \qquad (49)$$

where

$$\{...\} \equiv \left\langle \frac{\partial \mathbf{R}}{\partial g^{ik}} \right\rangle \delta g^{ik} - \partial_l \left[ \left\langle \frac{\partial \mathbf{R}}{\partial g_{lab}} \right\rangle - \partial_m \left\langle \frac{\partial \mathbf{R}}{\partial g_{lmab}} \right\rangle \right] \delta g_{ab} . \qquad (50)$$

At last substituting $\delta g_{ab} = -g_{ia} g_{kb} \delta g^{ik}$ according to relation (5) proves (45), since the variation

$$\mathbf{Q}^l \equiv \left[ \left\langle \frac{\partial \mathbf{R}}{\partial g_{lab}} \right\rangle - \partial_m \left\langle \frac{\partial \mathbf{R}}{\partial g_{lmab}} \right\rangle \right] \delta g_{ab} + \left\langle \frac{\partial \mathbf{R}}{\partial g_{lmab}} \right\rangle \delta g_{mab} \qquad (51)$$

vanishes at the boundary. – Concerning the question of priority in this context, already Corry, Renn, & Stachel [23], who emphasized the fact long before, that Hilbert [2] presented the tensor density (1) without calculation, concluded that he did not anticipate Einstein's definite form of the field equations. They dismissed his false argument, that the tensor density (1) had been the only thinkable result because of its constituents.

In view of the strange dating "20. November 1915" of Hilbert's [2] note – published in 1916 with a citation of Einstein's "25. November 1915" article [24], the latter actually containing the final results – Hilbert might have tacitly used Einstein's additional realization that the covariant derivative of the tensor $E_{ik}$ must vanish identically (equivalent to the contracted second *Bianchi identity* only if including Einstein's unambiguous trace term), which insight is necessary to complete his argumentation, but has not been explicitly mentioned there.

Furthermore it may be pointed to the simple fact, that there are two sides of Einstein's equations. As shown above, Hilbert would hardly have found their left hand side from (39) by any straightforward calculation at that time. For the right hand side, in addition, he offered some speculation insufficient to reach physical applicability in this form. There, in contrast, Einstein had set the phenomenological energy-momentum tensor of matter and pressure already years ago, which including his 'geodesic' equations of motion has successfully proven applicable even in various extreme situations again and again. Therefore, in spite of Hilbert's impressive mathematical contributions to GRT, nobody can seriously claim him to have derived Einstein's equations completely in his first note [2] or its draft at all.

### e) Conclusion

In its conventional form, the Euler-Lagrange formalism is inappropriate to handle GRT's principle of least action straightforwardly. This is basically shown in Section 'Concepts and notation' already and may well be the reason that Einstein's and Hilbert's formulae (27), (39) have been involved in some strange affairs of GRT history as addressed above. Since (36) is right – as explicitly verified in Section III of the supplementary material – expression (27) as it stands is apparently different from $\mathbf{E}_{ik}$, given one is taking into account, what partial derivatives $\partial(..)/\partial g^{ik}$, $\partial(..)/\partial g_l^{ik}$, $\partial(..)/\partial g_{lm}^{ik}$ mean when they are understood as well-defined single-component operations.

If – instead of the natural treatment straightforwardly leading to (33) – one insists on the assumed presupposition that the Lagrangian densities (28), (40) are functions of $g^{ik}$, $g_l^{ik}$, and $g_{lm}^{ik}$ only, then in the meaning of conventional tensor analysis the calculation seems to lead into a blind alley, since several operations like e.g. '$\partial g_{ab}/\partial g_{(lm)}^{ik}$' – unusual to the present in this form – are not even properly defined unambiguously there.[14]

Thus, concluding this note, the mathematical background of some unnecessary complications with (27), (39) may be summarized in the following statement: the treatment of a variational principle by applying the well-known conventional Euler-Lagrange equations in the framework of GRT would presuppose that (a) the total differential of the respective *action integral* is a true 4-dimensional scalar and (b) the respective *action density* is algebraically composed of quantities $f$ together with their derivatives $f_k \equiv \partial f/\partial x^k$ completely, where – contradicting the inequality below (3) – the usual differential $\delta f = f_k \delta x^k$ would directly apply. While in Einstein's early attempt [10] it has been primarily the presupposition (a) which was not fulfilled, then in his later fundamental 1916 articles [1] or [18] it is still presupposition (b) which seems not sufficiently taken into account as little as in Hilbert's treatment [2], too.

In contrast, both integral principles (31), (43) – if subsequently the variations are calculated to get the *total differentials* with help of *partial differentiations* according to (33), (36) or (45), (47) – do straightforwardly work.

On the one hand, it seems that the natural chance has been ignored to handle the action principle according to the most simple comparable derivations of (33), (45) if not directly according to e.g. Weyl [6], Palatini [21], Weinberg [7], Wald [5], or Landau & Lifschitz [4] (the latter with the reservation mentioned below). On the other hand, several authors tried to adhere schematically to the standard Euler-Lagrange formalism which proved unexpectedly problematic as can be seen from the historical (non-)treatment by Hilbert 1915/16 in [2], by Einstein[15] 1916 in [1], by Pauli 1921 in

---

[14] Moreover, if one tries to deal with pure tensor quantities, then after a replacement of $g_{lik} := -g_{ia} g_{kb} g_l^{ab}$ in $\mathbf{G}$ – necessary to apply (27) – any covariant tensor like e.g. $g_{ia}$ included there has to be dealt with as a constant. This, however, means that, if taken literally, expression like $\partial g_{kb}/\partial g^{ik}$ would even vanish temporarily.

[15] In §15 of [18] using the well-known coordinate constraint $\sqrt{\mathbf{g}} = 1$, he at first calculated from relation (47a) the total differential $\delta H$, then concluded in (49) what the partial derivatives should be, to come out in (47b) with the desired result at last. This treatment corresponds almost exactly to that of subsequently Landau & Lifschitz as mentioned in Footnote 12 above.

[3], and by Landau & Lifschitz 1992 in [4] {where the latter have also argued in favor of Einstein's formula (27)}. Otherwise, in view of various discussions and explicitly stated missing direct calculations of (27), (39), these authors would have probably mentioned that chance above.

Significantly Hilbert left out the calculation in [2] after he might have temporarily tried to evaluate formula (39) directly, thereby facing an effective impracticability if not proceeding in analogy to (47). Since a motivation to 'nostrify' Einstein's GRT by attempted prior publication of the gravitational equations appears more than likely in view of relevant correspondence (s. Doc.s 136, 139, 140, 144, 148, 149, 167 and especially Doc. 152 of CPAE Vol. **8**A), it seems hard to believe that he abstained from a direct calculation of his formula (39), if such a calculation had been a viable option within the standard treatment of tensor analysis at that time.

Particularly in view of his remarks on the numerical complication requiring an introduction of what has been called the 'Hilbert factors' above, it finally raises the question why Hilbert decided to apply the Euler-Lagrange formalism at all (*"... die n Lagrangeschen Variationsgleichungen"*), instead of focusing on the actual principle of least action without a usage of problematic partial derivatives, what soon afterwards Weyl [6] or Palatini [21] did alternatively.[16] This all the more since he apparently failed to complete the calculation necessary to claim a desired priority. The answer may be that, again, Hilbert proceeded on Einstein's tracks who already in his early 1914 attempt [10] had introduced the Euler-Lagrange formalism long before, whose conventional treatment is shown here not to apply directly.

### Acknowledgement

Author wishes to thank EPJH editors and referee for valuable criticism and continued skepticism. At last it has been Markus Fröb who after a public request via arXiv pointed out an inconsequent treatment of relation (II.A,2) in version v3, which now thanks to this hint has been fixed by introduction of a symbolic 'tensor differentiation' <..> in contrast to the straightforward 'component differentiation', overdue in the framework of general relativity. Though the existence of a 'strange detail' has been clear from the beginning in arXiv:gr-qc/0410068v1, it took several improvements and some corrections to reach this version v4 which except for potential minor changes or typos will be the final one. In view of a commonly misleading formal assumption and some missing treatments of authorities as in particular Einstein, Hilbert, Pauli, Landau & Lifschitz, such an updating procedure might be explicable after all. These authors once left something like a work in progress on a subtle puzzle concerning the unnecessary Euler-Lagrange approach to the superior principle of least action in the framework of GR without clear distinction between the conventional and a merely symbolic formalism explicitly defined now.

### Appendix

The original calculations of this appendix have been done to verify the relevant results in detail, at first still without a consistent visual distinction between various forms of differentiations. Here the completed calculations may be given in their final form as supplementary material appended to the electronic edition of this note.

---

[16] Probably the reason why something like Palatini's later concept [21] did not play a major role in Einstein's considerations, may have been that he regarded his fundamental tensor $g_{ik}$ the primary element instead of Christoffel's symbols $\Gamma^i_{kl}$ then. On the other hand, Palatini's method without explicitly using any partial derivatives $\partial/\partial g^{ik}$, $\partial/\partial g_l^{ik}$, $\partial/\partial g_{lm}^{ik}$ seems to have been developed in reaction to the unsatisfactory experiences of his mentor Levi-Civita in the discussion with Einstein addressed above (s. also [25]).

# Supplementary Material

for

"A Strange Detail
Concerning the Variational Principle
of General Relativity Theory"

## I. Some additional relations and calculation rules

$$G \equiv g^{um} G_{um} \tag{I,1}$$

$$G_{um} = g^{sv} g^{rw} C_{umsvrw} \tag{I,2}$$

$$C_{umsvrw} \equiv \{\Gamma_{w,ms}\Gamma_{v,ur} - \Gamma_{w,sr}\Gamma_{v,um}\} \tag{I,3}$$

$$\mathbf{G} = \sqrt{\mathbf{g}}\, g^{um} g^{sv} g^{rw} \{\Gamma_{v,ur}\Gamma_{w,ms} - \Gamma_{v,um}\Gamma_{w,sr}\} \tag{I,4}$$

$$g_{lik} \equiv \partial_l g_{ik} \equiv \partial g_{ik}/\partial x^l \tag{I,5}$$

$$g_l^{ik} \equiv \partial_l g^{ik} \equiv \partial g^{ik}/\partial x^l . \tag{I,6}$$

Several of the relations in this section may essentially coincide with others already given in the main text above. For a convenient cyclic indexing in Christoffel symbols (1. kind), the expression $g_{bac} \equiv \partial g_{ac}/\partial x^b$ may be replaced by $g_{bca} \equiv \partial g_{ca}/\partial x^b$ what is possible due to the $g_{ik}$-symmetry presupposed in this context:

$$\Gamma_{a,bc} \equiv \tfrac{1}{2}(g_{cab} + g_{bac} - g_{abc}) = \tfrac{1}{2}(g_{cab} + g_{bca} - g_{abc}) . \tag{I,7}$$

The negative determinant $-|g_{ik}| = \mathbf{g} = -1/|g^{ik}|$ is regarded a function of $g^{ik}$, corresponding to relation (24) above. Renaming ($m := u$, $r := s$, $v := w$) of indices or local exchange ($i \leftrightarrow k$), ($a \leftrightarrow b$) due to the respective symmetries of e.g. $g^{ik}$ or $g_{ab}$ where actually present will be repeatedly applied also below. Obviously it is

$$\left\langle \frac{\partial g^{ab}}{\partial g^{ik}} \right\rangle = \Delta^{ab}_{ik} \equiv \tfrac{1}{2}\left(\delta^a_i \delta^b_k + \delta^a_k \delta^b_i\right) \tag{I,8}$$

$$\left\langle \frac{\partial g_{ab}}{\partial g_{ik}} \right\rangle = \Delta^{ik}_{ab} \equiv \tfrac{1}{2}\left(\delta^i_a \delta^k_b + \delta^k_a \delta^i_b\right) \tag{I,9}$$

$$\left\langle \frac{\partial g^{ab}_c}{\partial g^{ik}_l} \right\rangle = \Delta^{lab}_{cik} = \tfrac{1}{2}\delta^l_c\left(\delta^a_i \delta^b_k + \delta^a_k \delta^b_i\right). \tag{I,10}$$

After partial derivations with respect to $g^{ik}$, any coherent expressions containing the indices $i$ and $k$ are tacitly understood to be symmetrized according to the explicit '$\Delta$-rules' above and below, leading to e.g. $g_{ai}g_{bk} = \tfrac{1}{2}(g_{ai}g_{bk}+g_{ak}g_{bi})$, where necessary. Those procedures apply to other partial derivations with respect to $g_{ab}, g_{lik} \ldots$ correspondingly. – In the rest of this section there are given some more relations, without further explanation completing 'Concepts and notation' above.

$$\left\langle \frac{\partial\sqrt{\mathbf{g}}}{\partial g^{ik}} \right\rangle \delta g^{ik} = \delta\sqrt{\mathbf{g}} = -\tfrac{1}{2}\sqrt{\mathbf{g}}\, g_{ik}\, \delta g^{ik} = \frac{1}{X}\frac{\partial\sqrt{\mathbf{g}}}{\partial g^{ik}}\delta g^{ik} . \tag{I,11}$$

$$\delta g_{ab} = -\tfrac{1}{2}\left(g_{ia}g_{kb} + g_{ka}g_{ib}\right)\delta g^{ik}$$
$$\delta g^{ik} = -\tfrac{1}{2}\left(g^{ia}g^{kb} + g^{ka}g^{ib}\right)\delta g_{ab} \tag{I,12}$$

$$g_{lik} = -\tfrac{1}{2}\left(g_{ia}g_{kb} + g_{ka}g_{ib}\right)g_l^{ab}$$
$$g_l^{ik} = -\tfrac{1}{2}\left(g^{ia}g^{kb} + g^{ka}g^{ib}\right)g_{lab} \tag{I,13}$$

$$\left\langle \frac{\partial[..]}{\partial g^{ik}_{(lm)}} \right\rangle \equiv \frac{1}{Y}\frac{\partial[..]}{\partial g^{ik}_{(lm)}}$$
$$\left\langle \frac{\partial[..]}{\partial g_{(lm)ik}} \right\rangle \equiv \frac{1}{Y}\frac{\partial[..]}{\partial g_{(lm)ik}} \tag{I,14}$$

$$\delta_{\mathrm{III}}[..] = \ldots + \left\langle \frac{\partial[..]}{\partial g^{ik}_{(lm)}} \right\rangle \delta g^{ik}_{(lm)} + \ldots$$
$$\delta_{\mathrm{IV}}[..] = \ldots + \left\langle \frac{\partial[..]}{\partial g_{(lm)ik}} \right\rangle \delta g_{(lm)ik} + \ldots \tag{I,15}$$

$$\delta g_c^{ab} = \Delta^{lab}_{cik}\,\delta g_l^{ik}$$
$$\left\langle \frac{\partial g_c^{ab}}{\partial g_l^{ik}} \right\rangle = \Delta^{lab}_{cik} = \frac{1}{X}\frac{\partial g_c^{ab}}{\partial g_l^{ik}}$$
$$\Delta^{lab}_{cik} \equiv \tfrac{1}{2}\delta^l_c\left(\delta^a_i\delta^b_k + \delta^a_k\delta^b_i\right) \tag{I,16}$$

$$\delta g_{cd}^{ab} = \Delta^{lmab}_{cdik}\,\delta g_{lm}^{ik}$$
$$\left\langle \frac{\partial g_{cd}^{ab}}{\partial g_{lm}^{ik}} \right\rangle = \Delta^{lmab}_{cdik} = \frac{1}{Y}\frac{\partial g_{cd}^{ab}}{\partial g_{lm}^{ik}}$$
$$\Delta^{lmab}_{cdik} \equiv \tfrac{1}{2}\left(\delta^l_c\delta^m_d + \delta^m_c\delta^l_d\right)\left(\delta^a_i\delta^b_k + \delta^a_k\delta^b_i\right) \tag{I,17}$$

$$\delta g_{ab} = \Delta^{ik}_{ab}\,\delta g_{ik}$$
$$\left\langle \frac{\partial g_{ab}}{\partial g_{ik}} \right\rangle = \Delta^{ik}_{ab} \equiv \frac{1}{X}\frac{\partial g_{ab}}{\partial g_{ik}}$$
$$\Delta^{ik}_{ab} \equiv \tfrac{1}{2}\left(\delta^i_a\delta^k_b + \delta^k_a\delta^i_b\right) \tag{I,18}$$

$$\delta g_{cdab} = \Delta^{lmik}_{cdab} \delta g_{lmik}$$

$$\left\langle \frac{\partial g_{cdab}}{\partial g_{lmik}} \right\rangle = \Delta^{lmik}_{cdab} = \frac{1}{Y} \frac{\partial g_{cdab}}{\partial g_{lmik}} \qquad (I,19)$$

$$\Delta^{lmik}_{cdab} \equiv \tfrac{1}{2} \left( \delta^l_c \delta^m_d + \delta^m_c \delta^l_d \right) \left( \delta^i_a \delta^k_b + \delta^k_a \delta^i_b \right)$$

As presupposed in the main text of the note, it is frequently used the possible exchange of differentiation with variation

$$\begin{aligned}\delta g^{ik}_l &= \partial_l \left( \delta g^{ik} \right) \\ \delta g_{lik} &= \partial_l \left( \delta g_{ik} \right)\end{aligned} \qquad (I,20)$$

An occasional treatment according to a temporarily assumed independence of $g^{ik}$ and $g_{ab}$, (s. in particular Section III) is justified by the fact that applying the chain rule of differentiation to any expression $E$, one is free to take it as composed of arbitrary constituents, as long as these do represent $E$ completely.

## II. Proof for the main equation (33) of the note

According to the main equation of the note, one claim is

$$\mathbf{E}_{ik} = \left\langle \frac{\partial \mathbf{G}}{\partial g^{ik}} \right\rangle + g_{ia} g_{kb} \partial_l \left\langle \frac{\partial \mathbf{G}}{\partial g_{lab}} \right\rangle . \tag{33}$$

According to (I,4), (I,7), evidently *G* may be regarded algebraically composed of the *contravariant* tensor $g^{ik}$ and the first derivatives $g_{lik}$ of the *covariant* tensor $g_{ik}$ only, which statement is already sufficient for the simple derivation of relation (33). According to the superior principle of least action, the total variation of the complete integral including the additive Lagrangian of matter has to vanish. Regarding the purely gravitational part, here it is

$$\delta \int \mathbf{G} \, d\Omega = \int \delta \mathbf{G} \, d\Omega = \int \left\{ \left\langle \frac{\partial \mathbf{G}}{\partial g^{ik}} \right\rangle \delta g^{ik} + \left\langle \frac{\partial \mathbf{G}}{\partial g_{lab}} \right\rangle \delta g_{lab} \right\} d\Omega \tag{II,1}$$

and by partial integration, after the exchange of differentiation with variation $\delta(\partial_l ..) = \partial_l (\delta ..)$ according to (I,20), the right hand side is found

$$\int \left\{ \left\langle \frac{\partial \mathbf{G}}{\partial g^{ik}} \right\rangle \delta g^{ik} - \partial_l \left\langle \frac{\partial \mathbf{G}}{\partial g_{lab}} \right\rangle \delta g_{ab} \right\} d\Omega \tag{II,2}$$

since as usual all variations may be presupposed to vanish at the boundaries of integration. Due to (5) then it follows

$$\delta \int \mathbf{G} \, d\Omega = \int \left[ \left\langle \frac{\partial \mathbf{G}}{\partial g^{ik}} \right\rangle + g_{ia} g_{kb} \partial_l \left\langle \frac{\partial \mathbf{G}}{\partial g_{lab}} \right\rangle \right] \delta g^{ik} \, d\Omega = 0 . \tag{II,3}$$

## II.A Calculation of $\left\langle \dfrac{\partial \mathbf{G}}{\partial g^{ik}} \right\rangle$

$$\left\langle \frac{\partial \mathbf{G}}{\partial g^{ik}} \right\rangle \equiv \tfrac{1}{2}(\mathbf{Z}_{ik} + \mathbf{Z}_{ki}) = \left\{ G_{ik} - \tfrac{1}{2} g_{ik} G \right\} + \mathbf{A}_{(ik)} \tag{II.A,1}$$

where $\mathbf{A}_{(ik)}$ means $\mathbf{A}_{ik}$ symmetrized with respect to the indices $i$, $k$, according to $\mathbf{A}_{(ik)} \equiv \tfrac{1}{2}(\mathbf{A}_{ik} + \mathbf{A}_{ki})$, where

$$\mathbf{A}_{ik} = \sqrt{\mathbf{g}}\, g^{um} g^{rw} \left[ 2\Gamma_{k,ur}\Gamma_{w,mi} - \Gamma_{k,um}\Gamma_{w,ir} - \Gamma_{w,um}\Gamma_{k,ri} \right] \tag{II.A,2}$$

This follows easily from (I,4), since

$$\frac{1}{\sqrt{\mathbf{g}}} \left\langle \frac{\partial \mathbf{G}}{\partial g^{ik}} \right\rangle = \frac{G}{\sqrt{\mathbf{g}}} \left\langle \frac{\partial \sqrt{\mathbf{g}}}{\partial g^{ik}} \right\rangle + \left\langle \frac{\partial G}{\partial g^{ik}} \right\rangle$$

$$=$$

$$-\tfrac{1}{2} g_{ik} G + \left( \Delta^{um}_{ik} g^{sv} g^{rw} + g^{um} \Delta^{sv}_{ik} g^{rw} + g^{um} g^{sv} \Delta^{rw}_{ik} \right) \left\{ \Gamma_{v,ur}\Gamma_{w,ms} - \Gamma_{v,um}\Gamma_{w,sr} \right\} \tag{II.A,3}$$

may be written

$$=$$

$$-\tfrac{1}{4} g_{ik} G + \tfrac{1}{2} \left[ g^{sv} g^{rw} \delta^u_i \delta^m_k + g^{um} g^{rw} \delta^s_i \delta^v_k + g^{um} g^{sv} \delta^r_i \delta^w_k \right] \left\{ \Gamma_{v,ur}\Gamma_{w,ms} - \Gamma_{v,um}\Gamma_{w,sr} \right\}$$

$$-\tfrac{1}{4} g_{ki} G + \tfrac{1}{2} \left[ g^{sv} g^{rw} \delta^u_k \delta^m_i + g^{um} g^{rw} \delta^s_k \delta^v_i + g^{um} g^{sv} \delta^r_k \delta^w_i \right] \left\{ \Gamma_{v,ur}\Gamma_{w,ms} - \Gamma_{v,um}\Gamma_{w,sr} \right\} \tag{II.A,4}$$

where the second line $\equiv \tfrac{1}{2} Z_{ki}$ is the same as the first line $\equiv \tfrac{1}{2} Z_{ik}$ with only the indices i, k exchanged. The evaluation now yields

$$Z_{ik}$$
$$=$$
$$-\tfrac{1}{2} g_{ik} G + g^{sv} g^{rw} \left[ \Gamma_{v,ir}\Gamma_{w,ks} - \Gamma_{v,ik}\Gamma_{w,sr} \right]$$
$$+ g^{um} g^{rw} \left[ \Gamma_{k,ur}\Gamma_{w,mi} - \Gamma_{k,um}\Gamma_{w,ir} \right] + g^{um} g^{sv} \left[ \Gamma_{v,ui}\Gamma_{k,ms} - \Gamma_{v,um}\Gamma_{k,si} \right]$$
$$=$$
$$G_{ik} - \tfrac{1}{2} g_{ik} G + g^{um} g^{rw} \left[ \Gamma_{k,ur}\Gamma_{w,mi} - \Gamma_{k,um}\Gamma_{w,ir} \right] + g^{um} g^{rw} \left[ \Gamma_{w,mi}\Gamma_{k,ur} - \Gamma_{w,um}\Gamma_{k,ri} \right]$$
$$=$$
$$G_{ik} - \tfrac{1}{2} g_{ik} G + g^{um} g^{rw} \left[ 2\Gamma_{k,ur}\Gamma_{w,mi} - \Gamma_{k,um}\Gamma_{w,ir} - \Gamma_{w,um}\Gamma_{k,ri} \right] \tag{II.A,5}$$

which result confirms (II.A,1), (II.A,2) rewriting it now by definition of $A_{ik}$ as

$$Z_{ik} \equiv G_{ik} - \tfrac{1}{2} g_{ik} G + A_{ik} \,. \tag{II.A,6}$$

## II.B Calculation of $g_{ia}g_{kb}\partial_l\left\langle\dfrac{\partial\mathbf{G}}{\partial g_{lab}}\right\rangle$

Since for partial derivations with respect to $g_{lab}$ obviously $\sqrt{\mathbf{g}}$ (as well as $g^{ik}$) has to be treated as constants, therefore

$$\left\langle\frac{\partial\mathbf{G}}{\partial g_{lab}}\right\rangle = \sqrt{\mathbf{g}}\left\langle\frac{\partial G}{\partial g_{lab}}\right\rangle \equiv \tfrac{1}{2}\sqrt{\mathbf{g}}\left(G^{lab}+G^{lba}\right) \qquad\text{(II.B,1)}$$

Now from (I,1), (I,2), (I,3) it is

$$\left\langle\frac{\partial G}{\partial g_{lab}}\right\rangle$$
$$=$$
$$g^{um}g^{sv}g^{rw}\left\langle\frac{\partial\{\Gamma_{v,ur}\Gamma_{w,ms}-\Gamma_{v,um}\Gamma_{w,sr}\}}{\partial g_{lab}}\right\rangle \qquad\text{(II.B,2)}$$

$$=$$

$$\tfrac{1}{2}\Gamma_{w,ms}\,g^{um}g^{sv}g^{rw}\left\langle\frac{\partial(g_{rvu}+g_{urv}-g_{vur})}{\partial g_{lab}}\right\rangle$$
$$+\tfrac{1}{2}\Gamma_{v,ur}\,g^{um}g^{sv}g^{rw}\left\langle\frac{\partial(g_{swm}+g_{msw}-g_{wms})}{\partial g_{lab}}\right\rangle$$
$$-\tfrac{1}{2}\Gamma_{w,sr}\,g^{um}g^{sv}g^{rw}\left\langle\frac{\partial(g_{mvu}+g_{umv}-g_{vum})}{\partial g_{lab}}\right\rangle \qquad\text{(II.B,3)}$$
$$-\tfrac{1}{2}\Gamma_{v,um}\,g^{um}g^{sv}g^{rw}\left\langle\frac{\partial(g_{rws}+g_{srw}-g_{wsr})}{\partial g_{lab}}\right\rangle$$

Correspondingly to the treatment in Section II.A, it is sufficient to calculate a non-symmetric $G^{lab}$ as part of (II.B,1) at first before doing the final (*ab*)-symmetrization according to relations (21), (22) subsequently

$$G^{lab}$$
$$=$$
$$\tfrac{1}{2}\Gamma_{w,ms}\left(g^{bm}g^{sa}g^{lw}+g^{lm}g^{sb}g^{aw}-g^{am}g^{sl}g^{bw}\right)$$
$$+\tfrac{1}{2}\Gamma_{v,ur}\left(g^{ub}g^{lv}g^{ra}+g^{ul}g^{av}g^{rb}-g^{ua}g^{bv}g^{rl}\right) \qquad\text{(II.B,4)}$$
$$-\tfrac{1}{2}\Gamma_{w,sr}\left(g^{bl}g^{sa}g^{rw}+g^{la}g^{sb}g^{rw}-g^{ab}g^{sl}g^{rw}\right)$$
$$-\tfrac{1}{2}\Gamma_{v,um}\left(g^{um}g^{bv}g^{la}+g^{um}g^{lv}g^{ab}-g^{um}g^{av}g^{bl}\right)$$

$$=$$

$$\tfrac{1}{2}\Gamma_{w,ms}\left(g^{bm}g^{sa}g^{lw}+g^{lm}g^{sb}g^{aw}-g^{am}g^{sl}g^{bw}\right)$$
$$+\tfrac{1}{2}\Gamma_{m,sw}\left(g^{sb}g^{lm}g^{wa}+g^{sl}g^{am}g^{wb}g-g^{sa}g^{bm}g^{wl}\right) \quad\text{(II.B,5)}$$
$$-\tfrac{1}{2}\Gamma_{w,sm}\left(2g^{bl}g^{sa}g^{mw}+0-g^{ab}g^{sl}g^{mw}\right)$$
$$-\tfrac{1}{2}\Gamma_{s,wm}\left(0+g^{wm}g^{ls}g^{ab}-0\right)$$

and with some more renaming of indices, taking into account also the (*ab*)-symmetry already temporarily

$$=$$

$$g^{bm}g^{sa}g^{lw}\left(\Gamma_{w,ms}+\Gamma_{s,wm}-\Gamma_{m,sw}\right)$$
$$-g^{bl}g^{sa}g^{mw}\Gamma_{w,sm}$$
$$+\tfrac{1}{2}g^{ab}g^{sl}g^{mw}\left(\Gamma_{w,sm}-\Gamma_{s,wm}\right)$$
$$=$$
$$g^{bm}g^{sa}\Gamma^{l}_{ms}+g^{lw}\left[g^{bm}\Gamma^{a}_{wm}-g^{sa}\Gamma^{b}_{sw}\right] \quad\text{(II.B,6)}$$
$$-g^{bl}g^{sa}g^{mw}\Gamma_{w,sm}$$
$$+\tfrac{1}{2}g^{ab}\left(g^{sl}\Gamma^{m}_{sm}-g^{mw}\Gamma^{l}_{wm}\right)$$
$$=$$
$$g^{sa}\left(g^{bm}\Gamma^{l}_{ms}-g^{bl}\Gamma^{m}_{sm}\right)+\tfrac{1}{2}g^{ab}\left(g^{sl}\Gamma^{m}_{sm}-g^{ms}\Gamma^{l}_{sm}\right)$$

one gets

$$G^{lab} \;=\; \left(g^{as}g^{bm}-\tfrac{1}{2}g^{ab}g^{sm}\right)\Gamma^{l}_{sm} - \left(g^{as}g^{bl}-\tfrac{1}{2}g^{ab}g^{ls}\right)\Gamma^{m}_{sm} \quad\text{(II.B,7)}$$

From this interim result one goes on with

$$g_{ia}g_{kb}\partial_l\left(\sqrt{\mathbf{g}}\,G^{lab}\right)$$
$$=$$
$$g_{ia}g_{kb}\left(\sqrt{\mathbf{g}}\,\partial_l G^{lab}+G^{lab}\partial_l\sqrt{\mathbf{g}}\right) \quad\text{(II.B,8)}$$
$$=$$
$$\sqrt{\mathbf{g}}\,g_{ia}g_{kb}\left(\partial_l G^{lab}+\tfrac{1}{2}g^{qr}g_{lqr}G^{lab}\right)$$

where with

$$\partial_l G^{lab} \;=\; \partial_l\left[\left(g^{as}g^{bm}-\tfrac{1}{2}g^{ab}g^{sm}\right)\Gamma^{l}_{sm} - \left(g^{as}g^{bl}-\tfrac{1}{2}g^{ab}g^{ls}\right)\Gamma^{m}_{sm}\right]$$
$$= \left[\left(g^{as}g^{bm}-\tfrac{1}{2}g^{ab}g^{sm}\right)\partial_l\Gamma^{l}_{sm} - \left(g^{as}g^{bl}-\tfrac{1}{2}g^{ab}g^{ls}\right)\partial_l\Gamma^{m}_{sm}\right] + C^{ab} \quad\text{(II.B,9)}$$

and

$$C^{ab} = \Gamma^l_{sm}\partial_l\left(g^{as}g^{bm} - \tfrac{1}{2}g^{ab}g^{sm}\right) - \Gamma^m_{sm}\partial_l\left(g^{as}g^{bl} - \tfrac{1}{2}g^{ab}g^{ls}\right). \qquad (\text{II.B},10)$$

one has

$$g_{ia}g_{kb}\left(\partial_l G^{lab} - C^{ab}\right)$$
$$=$$
$$g_{ia}g_{kb}\left[\left(g^{as}g^{bm} - \tfrac{1}{2}g^{ab}g^{sm}\right)\partial_l\Gamma^l_{sm} - \left(g^{as}g^{bl} - \tfrac{1}{2}g^{ab}g^{ls}\right)\partial_l\Gamma^m_{sm}\right]$$
$$=$$
$$\left(g_{ia}g_{kb}g^{as}g^{bm}\partial_l\Gamma^l_{sm} - \tfrac{1}{2}g_{ia}g_{kb}g^{ab}g^{sm}\partial_l\Gamma^l_{sm}\right) - \left(g_{ia}g_{kb}g^{as}g^{bl}\partial_l\Gamma^m_{sm} - \tfrac{1}{2}g_{ia}g_{kb}g^{ab}g^{ls}\partial_l\Gamma^m_{sm}\right)$$
$$= \qquad (\text{II.B},11)$$
$$\left(\partial_l\Gamma^l_{ik} - \tfrac{1}{2}g_{ki}g^{sm}\partial_l\Gamma^l_{sm}\right) - \left(\partial_k\Gamma^m_{im} - \tfrac{1}{2}g_{ki}g^{ls}\partial_l\Gamma^m_{sm}\right)$$
$$=$$
$$\partial_l\Gamma^l_{ik} - \partial_k\Gamma^m_{im} - \tfrac{1}{2}g_{ki}\left[g^{ms}\left(\partial_l\Gamma^l_{sm} - \partial_m\Gamma^l_{sl}\right)\right]$$
$$\equiv$$
$$r_{ik} - \tfrac{1}{2}g_{ki}r$$

From (II.B,8) - (II.B,11) now it is

$$g_{ia}g_{kb}\partial_l\left(\sqrt{g}\,G^{lab}\right) = \left\{\mathbf{r}_{ik} - \tfrac{1}{2}g_{ik}\mathbf{r}\right\} + \mathbf{B}_{ik} \qquad (\text{II.B},12)$$

where

$$\mathbf{B}_{ik} = \sqrt{g}\,g_{ia}g_{kb}\left[C^{ab} + D^{ab}\right] \qquad (\text{II.B},13)$$

with $C^{ab}$ of (II.B,10) results in

$$C^{ab}$$
$$=$$
$$g^{as}g^{bm}\left[g^{rt}\left(-2g_{lst}\Gamma^l_{mr} + \tfrac{1}{2}g_{lsm}\Gamma^l_{rt} + g_{rmt}\Gamma^l_{sl}\right) + \Gamma^r_{sm}\Gamma^l_{rl}\right] \qquad (\text{II.B},14)$$
$$+ \tfrac{1}{2}g^{ab}g^{sm}g^{rt}\left(g_{lst}\Gamma^l_{mr} - g_{rmt}\Gamma^l_{sl}\right)$$

as will be shown in (II.x,3) below, and

$$D^{ab} = \tfrac{1}{2}g^{qr}g_{lqr}G^{lab}$$
$$= \qquad . \qquad (\text{II.B},15)$$
$$\tfrac{1}{2}g^{qr}g_{lqr}\left[\left(g^{as}g^{bm} - \tfrac{1}{2}g^{ab}g^{sm}\right)\Gamma^l_{sm} - \left(g^{as}g^{bl} - \tfrac{1}{2}g^{ab}g^{ls}\right)\Gamma^m_{sm}\right]$$

Adjunct II(x):  Calculation of $C^{ab}$

$$C^{ab}$$
$$=$$
$$\Gamma^l_{sm}\partial_l\left(g^{as}g^{bm} - \tfrac{1}{2}g^{ab}g^{sm}\right) - \Gamma^m_{sm}\partial_l\left(g^{as}g^{bl} - \tfrac{1}{2}g^{ab}g^{ls}\right)$$
$$=$$
$$\Gamma^l_{sm}\left(2g^{as}g^{bm}_l - \tfrac{1}{2}g^{ab}_l g^{sm} - \tfrac{1}{2}g^{ab}g^{sm}_l\right) - \Gamma^m_{sm}\left(g^{as}_l g^{bl} + g^{as}g^{bl}_l - \tfrac{1}{2}g^{ab}_l g^{ls} - \tfrac{1}{2}g^{ab}g^{ls}_l\right)$$
$$=$$
$$\Gamma^l_{sm}\left(-2g^{as}g^{xb}g^{ym}g_{lxy} + \tfrac{1}{2}g^{sm}g^{xb}g^{ya}g_{lxy}\right)$$
$$- \Gamma^m_{sm}\left(-g^{bl}g^{xa}g^{ys}g_{lxy} - g^{as}g^{xb}g^{yl}g_{lxy} + \tfrac{1}{2}g^{ls}g^{xb}g^{ya}g_{lxy}\right)$$
$$+ \tfrac{1}{2}g^{ab}\left(g^{xs}g^{ym}g_{lxy}\Gamma^l_{sm} - g^{xl}g^{ys}g_{lxy}\Gamma^m_{sm}\right)$$
$$=$$
$$g^{as}g^{bm}g^{rt}\left[-2g_{lst}\Gamma^l_{mr} + \tfrac{1}{2}g_{lsm}\Gamma^l_{rt} + \left(g_{mst} - \tfrac{1}{2}g_{tms}\right)\Gamma^l_{rl}\right]$$
$$+ g^{as}g^{bm}g^{lt}g_{lmt}\Gamma^r_{sr}$$
$$+ \tfrac{1}{2}g^{ab}g^{sm}\left(g^{rt}g_{lts}\Gamma^l_{rm} - g^{lt}g_{ltm}\Gamma^r_{sr}\right)$$
$$=$$
$$g^{as}g^{bm}\left[-2g^{rt}g_{lst}\Gamma^l_{mr} + \tfrac{1}{2}g^{rt}g_{lsm}\Gamma^l_{rt} + g^{lt}g_{lmt}\Gamma^r_{sr} + g^{rt}\left(g_{mst} - \tfrac{1}{2}g_{tsm}\right)\Gamma^l_{rl}\right] \quad \text{(II.x,1)}$$
$$+ \tfrac{1}{2}g^{ab}g^{sm}\left(g^{rt}g_{lst}\Gamma^l_{mr} - g^{lt}g_{lmt}\Gamma^r_{sr}\right)$$
$$=$$
$$g^{as}g^{bm}\left[-2g^{rt}g_{lst}\Gamma^l_{mr} + \tfrac{1}{2}g^{rt}g_{lsm}\Gamma^l_{rt} + g^{lt}g_{lmt}\Gamma^r_{sr} + g^{rt}\left(g_{mst} - \tfrac{1}{2}g_{tsm}\right)\Gamma^l_{rl}\right] \quad \text{(II.x,2)}$$
$$+ \tfrac{1}{2}g^{ab}g^{sm}\left(g^{rt}g_{lst}\Gamma^l_{mr} - g^{lt}g_{lmt}\Gamma^r_{sr}\right)$$

$$C^{ab}$$
$$=$$
$$g^{as}g^{bm}\left[g^{rt}\left(-2g_{lst}\Gamma^l_{mr} + \tfrac{1}{2}g_{lsm}\Gamma^l_{rt} + g_{rmt}\Gamma^l_{sl}\right) + \Gamma^r_{sm}\Gamma^l_{rl}\right] \quad \text{(II.x,3)}$$
$$+ \tfrac{1}{2}g^{ab}g^{sm}g^{rt}\left(g_{lst}\Gamma^l_{mr} - g_{rmt}\Gamma^l_{sl}\right)$$

### Adjunct II(y): Calculation of ($A_{ik} + B_{ik}$)

$$\frac{1}{\sqrt{g}}(\mathbf{A}_{ik} + \mathbf{B}_{ik}) =$$

$$g^{um}g^{rw}\left[2\Gamma_{k,ur}\Gamma_{w,mi} - \Gamma_{k,um}\Gamma_{w,ir} - \Gamma_{w,um}\Gamma_{k,ri}\right]$$

$$+ g_{ia}g_{kb}\left\{\begin{array}{l} g^{as}g^{bm}\left[g^{rt}\left(-2g_{lst}\Gamma^l_{mr} + \tfrac{1}{2}g_{lsm}\Gamma^l_{rt} + g_{rmt}\Gamma^l_{sl}\right) + \Gamma^r_{sm}\Gamma^l_{rl}\right] \\ + \tfrac{1}{2}g^{ab}g^{sm}g^{rt}\left(g_{lst}\Gamma^l_{mr} - g_{rmt}\Gamma^l_{sl}\right) \end{array}\right\} \quad\quad \text{(II.y,1)}$$

$$+ g_{ia}g_{kb}\tfrac{1}{2}g^{qr}g_{lqr}\left[\left(g^{as}g^{bm} - \tfrac{1}{2}g^{ab}g^{sm}\right)\Gamma^l_{sm} - \left(g^{as}g^{bl} - \tfrac{1}{2}g^{ab}g^{ls}\right)\Gamma^m_{sm}\right]$$

$$----$$

$$\frac{1}{\sqrt{g}}(\mathbf{A}_{ik} + \mathbf{B}_{ik}) =$$

$$g^{um}g^{rw}\left[2\Gamma_{k,ur}\Gamma_{w,mi} - \Gamma_{k,um}\Gamma_{w,ir} - \Gamma_{w,um}\Gamma_{k,ri}\right]$$

$$+ \left[g^{rt}\left(-2g_{lit}\Gamma^l_{kr} + \tfrac{1}{2}g_{lik}\Gamma^l_{rt} + g_{rkt}\Gamma^l_{il}\right) + \Gamma^r_{ik}\Gamma^l_{rl}\right] + \tfrac{1}{2}g_{ik}g^{sm}g^{rt}\left(g_{lst}\Gamma^l_{mr} - g_{rmt}\Gamma^l_{sl}\right) \quad \text{(II.y,2)}$$

$$+ \tfrac{1}{2}g^{qr}g_{lqr}\left(\Gamma^l_{ik} - \tfrac{1}{2}g_{ik}g^{sm}\Gamma^l_{sm}\right) - \tfrac{1}{2}g^{qr}g_{kqr}\Gamma^m_{im} + \tfrac{1}{4}g_{ik}g^{ls}g^{qr}g_{lqr}\Gamma^m_{sm}$$

$$----$$

$$\frac{1}{\sqrt{g}}(\mathbf{A}_{ik} + \mathbf{B}_{ik}) =$$

$$g^{um}\left[2\Gamma_{k,ur}\Gamma^r_{mi} - \Gamma_{k,um}\Gamma^r_{ir} - \Gamma^r_{um}\Gamma_{k,ri}\right]$$

$$+ g^{rt}\left(-2g_{lit}\Gamma^l_{kr} + \tfrac{1}{2}g_{lik}\Gamma^l_{rt} + g_{rkt}\Gamma^l_{il}\right) + \Gamma^r_{ik}\Gamma^l_{rl} + \tfrac{1}{2}g^{qr}g_{lqr}\Gamma^l_{ik} - \tfrac{1}{2}g^{qr}g_{kqr}\Gamma^m_{im} \quad \text{(II.y,3)}$$

$$+ \tfrac{1}{2}g_{ik}\,g^{sm}g^{rt}\left[\left(g_{lst}\Gamma^l_{mr} - g_{rmt}\Gamma^l_{sl}\right) + \tfrac{1}{2}\left(g_{mtr}\Gamma^l_{sl} - g_{ltr}\Gamma^l_{sm}\right)\right]$$

$$----$$

$$\frac{1}{\sqrt{g}}(\mathbf{A}_{ik} + \mathbf{B}_{ik}) = \Gamma^r_{ik}\Gamma^l_{rl} +$$

$$+ g^{um}\left[2\Gamma_{k,ur}\Gamma^r_{mi} - \Gamma_{k,um}\Gamma^r_{ir} - \Gamma^r_{um}\Gamma_{k,ri}\right] \quad\quad \text{(II.y,4)}$$

$$+ g^{um}\left(-2g_{lim}\Gamma^l_{ku} + \tfrac{1}{2}g_{lik}\Gamma^l_{um} + g_{ukm}\Gamma^l_{il} + \tfrac{1}{2}g_{lum}\Gamma^l_{ik} - \tfrac{1}{2}g_{kum}\Gamma^l_{il}\right)$$

$$+ \tfrac{1}{2}g_{ik}g^{rt}\left[g^{sm}\left(g_{lst}\Gamma^l_{mr} - \tfrac{1}{2}g_{ltr}\Gamma^l_{sm}\right) - \Gamma^s_{tr}\Gamma^l_{sl}\right]$$

$$----$$

$$\frac{1}{\sqrt{g}}(\mathbf{A}_{ik} + \mathbf{B}_{ik}) = \Gamma^r_{ik}\Gamma^l_{rl} +$$

$$+ g^{um}\left[2\Gamma_{k,ul}\Gamma^l_{mi}\right] \quad\quad \text{(II.y,5)}$$

$$+ g^{um}\left(-2g_{lku}\Gamma^l_{im} + \tfrac{1}{2}g_{lum}\Gamma^l_{ik}\right)$$

$$+ \tfrac{1}{2}g_{ik}g^{rt}\left[g^{sm}g_{lst}\Gamma^l_{mr} - 2\Gamma^s_{tr}\Gamma^l_{sl}\right]$$

$$----$$

$$(\mathbf{A}_{ik} + \mathbf{B}_{ik}) = -2\left\{\mathbf{G}_{ik} - \tfrac{1}{2}g_{ik}\mathbf{G}\right\} \quad\quad \text{(II.y,6)}$$

## II.C Result of Section II

Since from (II.A,1) and (II.B,12) it is

$$\left\langle \frac{\partial \mathbf{G}}{\partial g^{ik}} \right\rangle + g_{ia} g_{kb} \partial_l \left\langle \frac{\partial \mathbf{G}}{\partial g_{lab}} \right\rangle = \left\{ \mathbf{G}_{ik} - \tfrac{1}{2} g_{ik} \mathbf{G} \right\} + \mathbf{A}_{(ik)} + \left\{ \mathbf{r}_{ik} - \tfrac{1}{2} g_{ik} \mathbf{r} \right\} + \mathbf{B}_{(ik)} \tag{II.C,1}$$

then with (II.y,6)

$$\left( \mathbf{A}_{ik} + \mathbf{B}_{ik} \right) = -2 \left\{ \mathbf{G}_{ik} - \tfrac{1}{2} g_{ik} \mathbf{G} \right\} \tag{II.C,2}$$

it finally results

$$\begin{aligned}
\left\langle \frac{\partial \mathbf{G}}{\partial g^{ik}} \right\rangle &+ g_{ia} g_{kb} \partial_l \left\langle \frac{\partial \mathbf{G}}{\partial g_{lab}} \right\rangle \\
&= \\
\mathbf{r}_{ik} &- \mathbf{G}_{ik} - \tfrac{1}{2} g_{ik} (\mathbf{r} - \mathbf{G}) \\
&\equiv \\
\mathbf{R}_{ik} &- \tfrac{1}{2} g_{ik} \mathbf{R} \\
&\equiv \\
&\mathbf{E}_{ik}
\end{aligned} \tag{II.C,3}$$

(q. e. d.)

## III. Proof for the additional equation (36) of the note

According to equation (36) of the note, one additional claim is

$$\mathbf{E}_{ik} = \left\langle \frac{\partial \mathbf{G}^{(ik)}}{\partial g^{ik}} \right\rangle - \partial_l \left\langle \frac{\partial \mathbf{G}}{\partial g_l^{ik}} \right\rangle - g_{ia} g_{kb} \left\langle \frac{\partial \mathbf{G}_{(ab)}}{\partial g_{ab}} \right\rangle \tag{36}$$

where as stated in the note $\partial \mathbf{G}^{(ik)}/\partial g^{ik}$ and $\partial \mathbf{G}_{(ab)}/\partial g_{ab}$ obviously mean independent derivations of the same Lagrangian $\mathbf{G}$ with respect to contravariant $g^{ik}$ and covariant $g_{ab}$ tensor factors appearing there now. As otherwise unnecessarily circumstantial functions of $g_l^{ik}$-quantities instead of simple $g_{lab}$-quantities in Section II here it is

$$\Gamma_{a,bc} = -\tfrac{1}{2}\left( g_{am} g_{bn} g_c^{mn} + g_{cm} g_{an} g_b^{mn} - g_{bm} g_{cn} g_a^{mn} \right) \tag{III,1}$$

where has been made use of relation (6). Any covariant fundamental tensor $g_{ik}$ in (I,7) – which is the inverse of $g^{ik}$ respectively – cannot be expressed as an algebraic function of the contravariant tensor $g^{ik}$ *as a whole* (though it can be expressed as a matrix of functions composed of 3- and 4-factor combinations of its single components). Now to deal with *G* as algebraically composed of tensors and their first derivatives only, one has to take $g^{ik}$ and $g_{ab}$ as temporarily independent quantities, just as presupposed to derive relation (36) in the note above.

Inserting $\Gamma_{a,bc}$ from (III,1) into (I,3), then according to (I,2) one finds *G* of (I,1) a function of $g^{ik}$, $g_l^{ik}$, *and* $g_{ab}$ now, where $C_{umsvrw}$, originally given by the curly brackets in (I,3), then obviously depends only on $g_l^{ik}$ and $g_{ab}$ at last. Making use of this, expression (28) may correspondingly be written

$$\mathbf{G} = \sqrt{\mathbf{g}}\, g^{um} g^{sv} g^{rw} C_{umsvrw} = \sqrt{\mathbf{g}}\, g^{um} g^{sv} g^{rw} \left\{ \Gamma_{w,ms} \Gamma_{v,ur} - \Gamma_{w,sr} \Gamma_{v,um} \right\}. \tag{III,2}$$

### III.A Calculation of $\left\langle \dfrac{\partial \mathbf{G}^{(ik)}}{\partial g^{ik}} \right\rangle$

Now starting from (III,2), the chain rule yields

$$\left\langle \frac{\partial \mathbf{G}^{(ik)}}{\partial g^{ik}} \right\rangle = g^{um}g^{sv}g^{rw}C_{umsvrw}\left\langle \frac{\partial \sqrt{\mathbf{g}}}{\partial g^{ik}} \right\rangle + \sqrt{\mathbf{g}}\, C_{umsvrw}\left\langle \frac{\partial (g^{um}g^{sv}g^{rw})}{\partial g^{ik}} \right\rangle + \sqrt{\mathbf{g}}\, g^{um}g^{sv}g^{rw}\left\langle \frac{\partial C_{umsvrw}}{\partial g^{ik}} \right\rangle \quad \text{(III.A,1)}$$

where here again the quantity $\mathbf{g} \equiv -1/|g^{ik}|$ is regarded a function of $g^{ik}$ only [and $g_{ik}$ on the right hand side of (24) may be understood merely the inverse tensor of $g^{ik}$ correspondingly]. Using (I,8) it is

$$\left\langle \frac{\partial (g^{um}g^{sv}g^{rw})}{\partial g^{ik}} \right\rangle = \left( \Delta^{um}_{ik}g^{sv}g^{rw} + g^{um}\Delta^{sv}_{ik}g^{rw} + g^{um}g^{sv}\Delta^{rw}_{ik} \right) \quad \text{(III.A,2)}$$

as well as

$$\left\langle \frac{\partial C_{umsvrw}}{\partial g^{ik}} \right\rangle = 0 \quad \text{(III.A,3)}$$

where the latter relation results from the independence of explicitly $g^{ik}$ there. – Applying the relations (24), (I,8), (III.A,2), (III.A,3) into (III.A,1) above, one finds

$$\left\langle \frac{\partial \mathbf{G}^{(ik)}}{\partial g^{ik}} \right\rangle$$
$$=$$
$$-\tfrac{1}{2} g_{ik}\mathbf{G} + \sqrt{\mathbf{g}}\left( C_{(ik)svrw}g^{sv}g^{rw} + C_{um(ik)rw}g^{um}g^{rw} + C_{umsv(ik)}g^{um}g^{sv} \right) \quad \text{(III.A,4)}$$
$$=$$
$$-\tfrac{1}{2} g_{ik}\mathbf{G} + \mathbf{G}_{ik} + \sqrt{\mathbf{g}}\left( C_{um(ik)rw}g^{um}g^{rw} + C_{umsv(ik)}g^{um}g^{sv} \right)$$

where indices $(ik)$ mean a subsequent symmetrizing according to (I,8). Also in any interim result of corresponding expressions the positions of the indices $i$ and $k$ can be exchanged where appropriate. From (III.A,4), taking this into account, now it follows corresponding to same assignment in (II.A,6)

$$\mathbf{Z}_{ik} =$$
$$= \mathbf{G}_{ik} - \tfrac{1}{2} g_{ik}\mathbf{G} + \sqrt{\mathbf{g}}\, g^{um}g^{rw}\{\Gamma_{w,mi}\Gamma_{k,ur} - \Gamma_{w,ir}\Gamma_{k,um}\} + \sqrt{\mathbf{g}}\, g^{um}g^{sv}\{\Gamma_{k,ms}\Gamma_{v,ui} - \Gamma_{k,si}\Gamma_{v,um}\}$$
$$= \mathbf{G}_{ik} - \tfrac{1}{2} g_{ik}\mathbf{G} + \sqrt{\mathbf{g}}\, g^{um}g^{pq}\{\Gamma_{q,mi}\Gamma_{k,up} - \Gamma_{q,ip}\Gamma_{k,um} + \Gamma_{k,mq}\Gamma_{p,ui} - \Gamma_{k,qi}\Gamma_{p,um}\} \quad \text{(III.A,5)}$$
$$= \mathbf{G}_{ik} - \tfrac{1}{2} g_{ik}\mathbf{G} + \mathbf{g}^{rs}\{2\Gamma^{p}_{si}\Gamma_{k,rp} - \Gamma^{p}_{ip}\Gamma_{k,rs} - \Gamma_{k,qi}\Gamma^{q}_{rs}\}$$

and at last

$$\left\langle \frac{\partial \mathbf{G}^{(ik)}}{\partial g^{ik}} \right\rangle = \tfrac{1}{2}(\mathbf{Z}_{ik} + \mathbf{Z}_{ki}) \quad \text{(III.A,6)}$$

(which interim result, of course, equals that of Section II.A above)

### III.B Calculation of $-g_{ia}g_{kb}\left\langle\dfrac{\partial \mathbf{G}_{(ab)}}{\partial g_{ab}}\right\rangle$

From (III,2) it follows

$$-g_{ia}g_{kb}\left\langle\frac{\partial \mathbf{G}_{(ab)}}{\partial g_{ab}}\right\rangle \equiv \tfrac{1}{2}(\mathbf{X}_{ik}+\mathbf{X}_{ki}) = -\sqrt{\mathbf{g}}\, g_{ia}g_{kb}\, g^{um}g^{sv}g^{rw}\left\langle\frac{\partial C_{umsvrw}}{\partial g_{ab}}\right\rangle \qquad \text{(III.B,1)}$$

where it has been taken into account that here $g^{ik}$ and $g_{ab}$ are temporarily independent quantities, and $\sqrt{\mathbf{g}}$ is treated as a function of the contravariant tensor $g^{ik}$ again (not of $g_{ab}$'s, what means $\partial \mathbf{G}/\partial g_{ab} = \sqrt{\mathbf{g}}\,\partial G/\partial g_{ab}$ since $\sqrt{\mathbf{g}}$ is regarded a constant with respect to $\partial(..)/\partial g_{ab}$ temporarily ). – Now using (I,9) and taken $\Gamma_{x,yz}$ according to (III,1), it is

$$\left\langle\frac{\partial \Gamma_{x,yz}}{\partial g_{ab}}\right\rangle$$

$$= -\tfrac{1}{2}\left\langle\frac{\partial\left(g_{xm}g_{yn}g_z^{mn}+g_{zm}g_{xn}g_y^{mn}-g_{ym}g_{zn}g_x^{mn}\right)}{\partial g_{ab}}\right\rangle \qquad \text{(III.B,2)}$$

$$= -\tfrac{1}{2}\left(\Delta^{ab}_{xm}g_{yn}g_z^{mn}+g_{xm}\Delta^{ab}_{yn}g_z^{mn}+\Delta^{ab}_{zm}g_{xn}g_y^{mn}+g_{zm}\Delta^{ab}_{xn}g_y^{mn}-\Delta^{ab}_{ym}g_{zn}g_x^{mn}-g_{ym}\Delta^{ab}_{zn}g_x^{mn}\right)$$

what means

$$\left\langle\frac{\partial \Gamma_{x,yz}}{\partial g_{ab}}\right\rangle \equiv \tfrac{1}{2}\left(Y_{abxyz}+Y_{baxyz}\right) \qquad \text{(III.B,3)}$$

where by definition

$$Y_{abxyz} = -\tfrac{1}{2}\left(\delta^a_x g_{yn}g_z^{bn}+g_{xm}\delta^a_y g_z^{mb}+\delta^a_z g_{xn}g_y^{bn}+g_{zm}\delta^a_x g_y^{mb}-\delta^a_y g_{zn}g_x^{bn}-g_{ym}\delta^a_z g_x^{mb}\right)$$

$$Y_{baxyz} = -\tfrac{1}{2}\left(\delta^b_x g_{yn}g_z^{an}+g_{xm}\delta^b_y g_z^{ma}+\delta^b_z g_{xn}g_y^{an}+g_{zm}\delta^b_x g_y^{ma}-\delta^b_y g_{zn}g_x^{an}-g_{ym}\delta^b_z g_x^{ma}\right) \qquad \text{(III.B,4)}$$

Also here, corresponding to the treatment in previous sections again, it is sufficient to go on with a non-symmetric $Y^{abxyz}$ as part of (III.B,4) at first, before doing the final $(ab)$-symmetrization according to (III.B,3) – or in general according to relation (I,9) – subsequently. Therefore it is

$$g_{ia}g_{kb}\,Y_{abxyz} =$$

$$= -\tfrac{1}{2}\left(g_{ix}g_{kb}g_{yn}g_z^{bn}+g_{iy}g_{kb}g_{xm}g_z^{mb}+g_{iz}g_{kb}g_{xn}g_y^{bn}+g_{ix}g_{kb}g_{zm}g_y^{mb}-g_{iy}g_{kb}g_{zn}g_x^{bn}-g_{iz}g_{kb}g_{ym}g_x^{mb}\right) \qquad \text{(III.B,5)}$$

$$= \tfrac{1}{2}\left(g_{ix}g_{zky}+g_{iy}g_{zxk}+g_{iz}g_{ykx}+g_{ix}g_{yzk}-g_{iy}g_{xkz}-g_{iz}g_{xyk}\right)$$

which interim result will be repeatedly used below. Coming back to (III.B,1), after insertion of (I,3) one finds

$$-g_{ia}g_{kb}\left\langle\frac{\partial \mathbf{G}_{(ab)}}{\partial g_{ab}}\right\rangle =$$

$$= -g_{ia}g_{kb}g^{um}g^{sv}g^{rw}\left\langle\frac{\partial\{\Gamma_{w,ms}\Gamma_{v,ur} - \Gamma_{w,sr}\Gamma_{v,um}\}}{\partial g_{ab}}\right\rangle \qquad \text{(III.B,6)}$$

$$= -g_{ia}g_{kb}g^{um}g^{sv}g^{rw}\left[\Gamma_{v,ur}\left\langle\frac{\partial\Gamma_{w,ms}}{\partial g_{ab}}\right\rangle + \Gamma_{w,ms}\left\langle\frac{\partial\Gamma_{v,ur}}{\partial g_{ab}}\right\rangle - \Gamma_{v,um}\left\langle\frac{\partial\Gamma_{w,sr}}{\partial g_{ab}}\right\rangle - \Gamma_{w,sr}\left\langle\frac{\partial\Gamma_{v,um}}{\partial g_{ab}}\right\rangle\right]$$

where after exchange of indices ($m \leftrightarrow u$, $r \leftrightarrow s$, $v \leftrightarrow w$) and using the $(ab)$-symmetry here and below, relation

$$g^{um}g^{sv}g^{rw}\Gamma_{w,ms}\left\langle\frac{\partial\Gamma_{v,ur}}{\partial g_{ab}}\right\rangle = g^{mu}g^{rw}g^{sv}\Gamma_{v,ur}\left\langle\frac{\partial\Gamma_{w,ms}}{\partial g_{ab}}\right\rangle \qquad \text{(III.B,7)}$$

means

$$-g_{ia}g_{kb}\left\langle\frac{\partial \mathbf{G}_{(ab)}}{\partial g_{ab}}\right\rangle =$$

$$= g^{um}g^{sv}g^{rw}\left[-2g_{ia}g_{kb}\Gamma_{v,ur}\left\langle\frac{\partial\Gamma_{w,ms}}{\partial g_{ab}}\right\rangle + g_{ia}g_{kb}\Gamma_{v,um}\left\langle\frac{\partial\Gamma_{w,sr}}{\partial g_{ab}}\right\rangle + g_{ia}g_{kb}\Gamma_{w,sr}\left\langle\frac{\partial\Gamma_{v,um}}{\partial g_{ab}}\right\rangle\right] \qquad \text{(III.B,8)}$$

and proceed making use of (III.B,5) to calculate the not yet symmetrized part $\mathbf{X}_{ik}$ of (III.B,1)

$$\mathbf{X}_{ik} =$$

$$\tfrac{1}{2}g^{um}g^{sv}g^{rw}\begin{bmatrix}-2\Gamma_{v,ur}(g_{iw}g_{skm} + g_{im}g_{swk} + g_{is}g_{mkw} + g_{iw}g_{msk} - g_{im}g_{wks} - g_{is}g_{wmk})\\ +\Gamma_{v,um}(g_{iw}g_{rks} + g_{is}g_{rwk} + g_{ir}g_{skw} + g_{iw}g_{srk} - g_{is}g_{wkr} - g_{ir}g_{wsk})\\ +\Gamma_{w,sr}(g_{iv}g_{mku} + g_{iu}g_{mvk} + g_{im}g_{ukv} + g_{iv}g_{umk} - g_{iu}g_{vkm} - g_{im}g_{vuk})\end{bmatrix}$$

$$= g^{um}g^{sv}g^{rw}\begin{bmatrix}-\Gamma_{v,ur}(g_{iw}g_{skm} + g_{im}g_{swk} + g_{is}g_{mkw} + g_{iw}g_{msk} - g_{im}g_{wks} - g_{is}g_{wmk})\\ +\tfrac{1}{2}\Gamma_{v,um}(g_{iw}g_{rks} + g_{is}g_{rwk} + g_{ir}g_{skw} + g_{iw}g_{srk} - g_{is}g_{wkr} - g_{ir}g_{wsk})\\ +\Gamma_{w,sr}(g_{iv}g_{mku} + g_{iu}g_{mvk} - g_{iu}g_{vkm})\end{bmatrix}$$

$$= \begin{bmatrix}-\Gamma^{s}_{ur}g^{um}g^{rw}(g_{iw}g_{skm} + g_{im}g_{swk} + g_{is}g_{mkw} + g_{iw}g_{msk} - g_{im}g_{wks} - g_{is}g_{wmk})\\ +\tfrac{1}{2}\Gamma^{s}_{um}g^{um}(g_{iks} + g^{rw}g_{is}g_{rwk} + g_{ski} + g_{sik} - g^{rw}g_{is}g_{wkr} - g_{isk})\\ +g^{sv}\Gamma^{r}_{sr}(g^{um}g_{iv}g_{mku} + g_{ivk} - g_{vki})\end{bmatrix} \qquad \text{(III.B,9)}$$

$$= \begin{bmatrix} \left(-\Gamma^s_{ui}g^{um}g_{skm} - \Gamma^s_{ir}g^{rw}g_{swk} - \Gamma^s_{ur}g^{um}g^{rw}g_{is}g_{mkw} - \Gamma^s_{ui}g^{um}g_{msk} + \Gamma^s_{ir}g^{rw}g_{wks} + \Gamma^s_{ur}g^{um}g^{rw}g_{is}g_{wmk}\right) \\ +g^{um}\Gamma^s_{um}g_{sik} \\ +g^{um}\Gamma^r_{ur}\left(g^{rs}g_{im}g_{skr} + g_{imk} - g_{mki}\right) \end{bmatrix}$$

$$= g^{um}\begin{bmatrix} -2\Gamma^s_{ui}g_{skm} + \Gamma^s_{ur}g^{rw}g_{is}\left(g_{wmk} - g_{mkw}\right) \\ +\Gamma^s_{um}g_{sik} \\ +\Gamma^r_{ur}\left(g^{rs}g_{im}g_{skr} + g_{imk} - g_{mki}\right) \end{bmatrix} \qquad \text{(III.B,10)}$$

The result of (III.B,1) is found by the final (*ik*)-symmetrization implied there, now from

$$\mathbf{X}_{ik} = \sqrt{\mathbf{g}}\, g^{rs}\left[-2g_{wks}\Gamma^w_{ri} + g^{pu}g_{iw}(g_{urk} - g_{rku})\Gamma^w_{sp} + g_{lik}\Gamma^l_{rs} + \left(g^{xy}g_{ir}g_{ykx} + g_{irk} - g_{rki}\right)\Gamma^p_{sp}\right] \qquad \text{(III.B,11)}$$

### III.C Calculation of $-\partial_l \left\langle \dfrac{\partial \mathbf{G}}{\partial g_l^{ik}} \right\rangle$

Since for partial derivations with respect to $g_l^{ik}$ obviously $\sqrt{\mathbf{g}}$ as well as $g^{ik}$, $g_{ab}$ have to be treated as temporary constants, it is

$$\left\langle \frac{\partial \mathbf{G}}{\partial g_l^{ik}} \right\rangle = \sqrt{\mathbf{g}} \left\langle \frac{\partial G}{\partial g_l^{ik}} \right\rangle \equiv \tfrac{1}{2}\sqrt{\mathbf{g}}\left(G_{ik}^l + G_{ki}^l\right) \tag{III.C,1}$$

From (28), (I,4) it is

$$G = g^{um}g^{sv}g^{rw}\left\{\Gamma_{w,ms}\Gamma_{v,ur} - \Gamma_{w,sr}\Gamma_{v,um}\right\} \tag{III.C,2}$$

and therefore

$$\left\langle \frac{\partial G}{\partial g_l^{ik}} \right\rangle = g^{um}g^{sv}g^{rw}\left\{ \Gamma_{v,ur}\left\langle\frac{\partial \Gamma_{w,ms}}{\partial g_l^{ik}}\right\rangle + \Gamma_{w,ms}\left\langle\frac{\partial \Gamma_{v,ur}}{\partial g_l^{ik}}\right\rangle - \Gamma_{v,um}\left\langle\frac{\partial \Gamma_{w,sr}}{\partial g_l^{ik}}\right\rangle - \Gamma_{w,sr}\left\langle\frac{\partial \Gamma_{v,um}}{\partial g_l^{ik}}\right\rangle \right\} \tag{III.C,3}$$

Now according to (I,10), from (III,1) one finds

$$\left\langle \frac{\partial \Gamma_{a,bc}}{\partial g_l^{ik}} \right\rangle = -\tfrac{1}{2}\left(g_{am}g_{bn}\Delta_{cik}^{lmn} + g_{cm}g_{an}\Delta_{bik}^{lmn} - g_{bm}g_{cn}\Delta_{aik}^{lmn}\right) \tag{III.C,4}$$

or preliminarily without explicit $(ik)$-symmetrization yet

$$\left\langle \frac{\partial \Gamma_{a,bc}}{\partial g_l^{ik}} \right\rangle_{\text{temp}} = -\tfrac{1}{2}\left(g_{ai}g_{bk}\delta_c^l + g_{ci}g_{ak}\delta_b^l - g_{bi}g_{ck}\delta_a^l\right) \tag{III.C,5}$$

With respect to the identity in (III.C,1), now from (III.C,3) and adapting (III.C,5) one finds

$$\begin{aligned}
G_{ik}^l &= \\
&-\tfrac{1}{2}g^{um}g^{sv}g^{rw}\Gamma_{v,ur}\left(g_{wi}g_{mk}\delta_s^l + g_{si}g_{wk}\delta_m^l - g_{mi}g_{sk}\delta_w^l\right) \\
&-\tfrac{1}{2}g^{um}g^{sv}g^{rw}\Gamma_{w,ms}\left(g_{vi}g_{uk}\delta_r^l + g_{ri}g_{vk}\delta_u^l - g_{ui}g_{rk}\delta_v^l\right) \\
&+\tfrac{1}{2}g^{um}g^{sv}g^{rw}\Gamma_{v,um}\left(g_{wi}g_{sk}\delta_r^l + g_{ri}g_{wk}\delta_s^l - g_{si}g_{rk}\delta_w^l\right) \\
&+\tfrac{1}{2}g^{um}g^{sv}g^{rw}\Gamma_{w,sr}\left(g_{vi}g_{uk}\delta_m^l + g_{mi}g_{vk}\delta_u^l - g_{ui}g_{mk}\delta_v^l\right)
\end{aligned} \tag{III.C,6}$$

$$=$$

$$-\tfrac{1}{2}\left(g_{wi}g_{mk}\delta^l_s g^{um}g^{sv}g^{rw}\Gamma_{v,ur}+g_{si}g_{wk}\delta^l_m g^{um}g^{sv}g^{rw}\Gamma_{v,ur}-g_{mi}g_{sk}\delta^l_w g^{um}g^{sv}g^{rw}\Gamma_{v,ur}\right)$$
$$-\tfrac{1}{2}\left(g_{vi}g_{uk}\delta^l_r g^{um}g^{sv}g^{rw}\Gamma_{w,ms}+g_{ri}g_{vk}\delta^l_u g^{um}g^{sv}g^{rw}\Gamma_{w,ms}-g_{ui}g_{rk}\delta^l_v g^{um}g^{sv}g^{rw}\Gamma_{w,ms}\right) \qquad \text{(III.C,7)}$$
$$+\tfrac{1}{2}\left(g_{wi}g_{sk}\delta^l_r g^{um}g^{sv}g^{rw}\Gamma_{v,um}+g_{ri}g_{wk}\delta^l_s g^{um}g^{sv}g^{rw}\Gamma_{v,um}-g_{si}g_{rk}\delta^l_w g^{um}g^{sv}g^{rw}\Gamma_{v,um}\right)$$
$$+\tfrac{1}{2}\left(g_{vi}g_{uk}\delta^l_m g^{um}g^{sv}g^{rw}\Gamma_{w,sr}+g_{mi}g_{vk}\delta^l_u g^{um}g^{sv}g^{rw}\Gamma_{w,sr}-g_{ui}g_{mk}\delta^l_v g^{um}g^{sv}g^{rw}\Gamma_{w,sr}\right)$$

$$=$$

$$-\tfrac{1}{2}\left(g^{lv}\Gamma_{v,ki}+g^{ul}\Gamma_{i,uk}-g^{rl}\Gamma_{k,ir}\right)$$
$$-\tfrac{1}{2}\left(g^{lw}\Gamma_{w,ki}+g^{lm}\Gamma_{i,mk}-g^{sl}\Gamma_{k,is}\right) \qquad \text{(III.C,8)}$$
$$+\tfrac{1}{2}\left(\delta^l_i g^{um}\Gamma_{k,um}+g_{ik}g^{um}g^{lv}\Gamma_{v,um}-\delta^l_k g^{um}\Gamma_{i,um}\right)$$
$$+\tfrac{1}{2}\left(\delta^l_k g^{rw}\Gamma_{w,ir}+\delta^l_i g^{rw}\Gamma_{w,kr}-g_{ik}g^{sl}g^{rw}\Gamma_{w,sr}\right)$$

$$=$$

$$-\tfrac{1}{2}\Gamma^l_{ki}-\tfrac{1}{2}g^{rl}\left(\Gamma_{i,kr}-\Gamma_{k,ir}\right)$$
$$-\tfrac{1}{2}\Gamma^l_{ki}-\tfrac{1}{2}g^{ls}\left(\Gamma_{i,ks}-\Gamma_{k,is}\right) \qquad \text{(III.C,9)}$$
$$+\tfrac{1}{2}g_{ik}g^{rs}\Gamma^l_{rs}+\tfrac{1}{2}g^{um}\left(\delta^l_i \Gamma_{k,um}-\delta^l_k \Gamma_{i,um}\right)$$
$$-\tfrac{1}{2}g_{ik}g^{ls}\Gamma^r_{sr}+\tfrac{1}{2}\left(\delta^l_k \Gamma^r_{ir}+\delta^l_i \Gamma^r_{kr}\right)$$

$$=$$

$$-\tfrac{1}{2}\Gamma^l_{ki}$$
$$-\tfrac{1}{2}\Gamma^l_{ki} \qquad \text{(III.C,10)}$$
$$+\tfrac{1}{2}g_{ik}g^{rs}\Gamma^l_{rs}$$
$$-\tfrac{1}{2}g_{ik}g^{ls}\Gamma^r_{sr}+\tfrac{1}{2}\left(\delta^l_k \Gamma^r_{ir}+\delta^l_i \Gamma^r_{kr}\right)$$

and thus

$$G^l_{ik} \equiv \frac{\partial G}{\partial g^{ik}_l} \;=\; -\Gamma^l_{ki}+\tfrac{1}{2}\left(\delta^l_k \Gamma^r_{ir}+\delta^l_i \Gamma^r_{kr}\right)+\tfrac{1}{2}g_{ik}\left(g^{rs}\Gamma^l_{rs}-g^{ls}\Gamma^r_{sr}\right) \qquad \text{(III.C,11)}$$

This interim result will be used now evaluating

$$\partial_l\left(\sqrt{\mathbf{g}}\,G^l_{ik}\right) = \sqrt{\mathbf{g}}\left\{\partial_l G^l_{ik} + G^l_{ik}\Gamma^m_{lm}\right\}$$
$$=$$
$$\sqrt{\mathbf{g}}\left[-\partial_l\Gamma^l_{ik} + \tfrac{1}{2}\left(\partial_k\Gamma^r_{ir} + \partial_i\Gamma^r_{kr}\right) + \tfrac{1}{2}g_{lik}\left(g^{rs}\Gamma^l_{rs} - g^{ls}\Gamma^r_{sr}\right) + \tfrac{1}{2}g_{ik}\partial_l\left(g^{rs}\Gamma^l_{rs} - g^{ls}\Gamma^r_{sr}\right)\right] \quad \text{(III.C,12)}$$
$$+\sqrt{\mathbf{g}}\left[-\Gamma^l_{ik}\Gamma^s_{ls} + \Gamma^s_{is}\Gamma^r_{kr} + \tfrac{1}{2}g_{ik}g^{ls}\left(\Gamma^r_{ls}\Gamma^m_{rm} - \Gamma^r_{sr}\Gamma^m_{lm}\right)\right]$$

what means

$$-\partial_l\left(\sqrt{\mathbf{g}}\,G^l_{ik}\right) =$$
$$\sqrt{\mathbf{g}}\left\{\partial_l\Gamma^l_{ik} - \partial_i\Gamma^r_{kr} + \Gamma^l_{ik}\Gamma^s_{ls} - \Gamma^s_{is}\Gamma^r_{kr} - \tfrac{1}{2}g_{ik}\left[\partial_l\left(g^{rs}\Gamma^l_{rs} - g^{ls}\Gamma^r_{sr}\right) + g^{rs}g^{al}\left(g_{sra} - g_{ars}\right)\Gamma^m_{lm}\right]\right\} \quad \text{(III.C,13)}$$
$$-\sqrt{\mathbf{g}}\,\Gamma_{k,li}\left(g^{rs}\Gamma^l_{rs} - g^{ls}\Gamma^r_{sr}\right)$$

what according to (III.C,1) has to be symmetrized at last by making use of

$$-\partial_l\left\langle\frac{\partial\mathbf{G}}{\partial g^{ik}_l}\right\rangle = -\partial_l\left[\sqrt{\mathbf{g}}\left\langle\frac{\partial G}{\partial g^{ik}_l}\right\rangle\right] \equiv -\tfrac{1}{2}\left[\partial_l\left(\sqrt{\mathbf{g}}\,G^l_{ik}\right) + \partial_l\left(\sqrt{\mathbf{g}}\,G^l_{ki}\right)\right]. \quad \text{(III.C,14)}$$

### III.D Result of Section III

Now one has

$$E_{ik} = \frac{1}{\sqrt{\mathbf{g}}} \left\{ \left\langle \frac{\partial \mathbf{G}^{(ik)}}{\partial g^{ik}} \right\rangle - \partial_l \left\langle \frac{\partial \mathbf{G}}{\partial g_l^{ik}} \right\rangle - g_{ia} g_{kb} \left\langle \frac{\partial \mathbf{G}_{(ab)}}{\partial g_{ab}} \right\rangle \right\} \tag{III.D,1}$$

using the non-symmetric interim results (III.A,5), (III.B,11), (III.C,13)

$$E_{ik} = \frac{1}{\sqrt{\mathbf{g}}} \left\{ \mathbf{Z}_{ik} - \partial_l \left( \sqrt{\mathbf{g}}\, G_{ik}^l \right) + \mathbf{X}_{ik} \right\}_{\text{symmetrized}} \tag{III.D,2}$$

which may be afterwards symmetrized now by exchange of the indices $(i \leftrightarrow k)$ where appropriate in the following. Thus

$$E_{ik}$$
$$=$$
$$\partial_l \Gamma_{ik}^l - \partial_i \Gamma_{kr}^r + \Gamma_{ik}^l \Gamma_{ls}^s - \Gamma_{is}^s \Gamma_{kr}^r - \tfrac{1}{2} g_{ik} \left[ \partial_l \left( g^{rs} \Gamma_{rs}^l - g^{ls} \Gamma_{sr}^r \right) + g^{rs} g^{al} (g_{sra} - g_{ars}) \Gamma_{lm}^m \right]$$
$$- \Gamma_{k,li} \left( g^{rs} \Gamma_{rs}^l - g^{ls} \Gamma_{sr}^r \right)$$
$$+ G_{ik} - \tfrac{1}{2} g_{ik} G + g^{rs} \left\{ 2 \Gamma_{si}^p \Gamma_{k,rp} - \Gamma_{ip}^p \Gamma_{k,rs} - \Gamma_{k,qi} \Gamma_{rs}^q \right\}$$
$$+ g^{rs} \left[ -2 g_{wks} \Gamma_{ri}^w + g^{pu} g_{iw} (g_{urk} - g_{rku}) \Gamma_{sp}^w + g_{lik} \Gamma_{rs}^l + \left( g^{xy} g_{ir} g_{ykx} + g_{irk} - g_{rki} \right) \Gamma_{sp}^p \right] \tag{III.D,3}$$
$$=$$
$$R_{ik} + \Gamma_{is}^r \Gamma_{kr}^s - \Gamma_{is}^s \Gamma_{kr}^r - \tfrac{1}{2} g_{ik} \left[ R + 2G + \left( g_l^{rs} \Gamma_{rs}^l - g_l^{ls} \Gamma_{sr}^r \right) + g^{rs} g^{al} (g_{sra} - g_{ars}) \Gamma_{lm}^m \right]$$
$$+ \Gamma_{is}^r \Gamma_{kr}^s - \tfrac{1}{2} g_{lik} \left( g^{ls} \Gamma_{sr}^r \right) - \Gamma_{ik}^s \Gamma_{sr}^r$$
$$+ g^{rs} \left\{ 2 \Gamma_{si}^l \Gamma_{k,rl} - \Gamma_{il}^l \Gamma_{k,rs} \right\}$$
$$+ g^{rs} \left[ -2 g_{wks} \Gamma_{ri}^w + g^{pu} g_{iw} (g_{urk} - g_{rku}) \Gamma_{sp}^w + \left( g^{xy} g_{ir} g_{ykx} + g_{irk} \right) \Gamma_{sp}^p \right]$$
$$=$$
$$R_{ik} + \Gamma_{is}^r \Gamma_{kr}^s - \Gamma_{is}^s \Gamma_{kr}^r - \tfrac{1}{2} g_{ik} \left[ G + \partial_l \left( g^{rs} \Gamma_{rs}^l - g^{ls} \Gamma_{sr}^r \right) + g^{rs} g^{al} (g_{sra} - g_{ars}) \Gamma_{lm}^m \right]$$
$$- \tfrac{1}{2} g_{lik} \left( g^{rs} \Gamma_{rs}^l - g^{ls} \Gamma_{sr}^r \right)$$
$$+ G_{ik} - \tfrac{1}{2} g_{lik} g^{rs} \Gamma_{rs}^l + g^{rs} \left\{ 2 \Gamma_{si}^p \Gamma_{k,rp} - \Gamma_{il}^l \Gamma_{k,rs} \right\}$$
$$+ g_{lik} g^{rs} \Gamma_{rs}^l + g^{rs} \left[ -2 g_{wks} \Gamma_{ri}^w + g^{pu} g_{iw} (g_{urk} - g_{rku}) \Gamma_{sp}^w + \left( g^{xy} g_{ir} g_{ykx} + g_{irk} - g_{rki} \right) \Gamma_{sp}^p \right] \tag{III.D,4}$$

$$=$$

$$R_{ik} + \Gamma^r_{is}\Gamma^s_{kr} - \Gamma^s_{is}\Gamma^r_{kr} - \tfrac{1}{2} g_{ik}\left[G + \left(g^{rs}\partial_l \Gamma^l_{rs} - g^{ls}\partial_l \Gamma^r_{sr}\right) + \left(g^{rs}_l \Gamma^l_{rs} - g^{ls}_l \Gamma^r_{sr}\right) + g^{rs}g^{al}(g_{sra} - g_{ars})\Gamma^m_{lm}\right]$$

$$+ \tfrac{1}{2} g_{lik}\left(g^{ls}\Gamma^r_{sr}\right) - g_{rki} g^{rs} \Gamma^p_{sp} + G_{ik}$$

$$+ g^{rs}\left\{ 2\Gamma^p_{si}\Gamma_{k,rp} - \Gamma^l_{il}\Gamma_{k,rs}\right\}$$

$$+ g^{rs}\left[-2g_{wks}\Gamma^w_{ri} + g^{pu}g_{iw}(g_{urk} - g_{rku})\Gamma^w_{sp} + \left(g^{xy}g_{ir}g_{ykx} + g_{irk}\right)\Gamma^p_{sp}\right]$$

$$=$$

$$R_{ik} + \Gamma^r_{is}\Gamma^s_{kr} - \Gamma^s_{is}\Gamma^r_{kr} - \tfrac{1}{2} g_{ik}\left[R + 2G + \left(g^{rs}_l \Gamma^l_{rs} - g^{ls}_l \Gamma^r_{sr}\right) + g^{rs}g^{al}(g_{sra} - g_{ars})\Gamma^m_{lm}\right]$$

$$+ \Gamma^r_{is}\Gamma^s_{kr} - g^{ls}(g_{kli})\Gamma^m_{sm}$$

$$+ g^{rs}\left\{ 2\Gamma^l_{si}\Gamma_{k,rl} - \Gamma^l_{il}\Gamma_{k,rs}\right\}$$

$$+ g^{rs}\left[-2g_{wks}\Gamma^w_{ri} + g^{pu}g_{iw}(g_{urk} - g_{rku})\Gamma^w_{sp} + \left(g^{xy}g_{ir}g_{ykx} + g_{irk}\right)\Gamma^p_{sp}\right]$$

$$=$$

$$R_{ik} - \tfrac{1}{2} g_{ik}\left[R + 2G + \left(g^{rs}_l \Gamma^l_{rs} - g^{ls}_l \Gamma^r_{sr}\right) + g^{rs}g^{al}(g_{sra} - g_{ars})\Gamma^m_{lm}\right]$$

$$+ 2\Gamma^r_{is}\Gamma^s_{kr} - \Gamma^s_{is}\Gamma^r_{kr} + \left[-g^{ls}(g_{kli})\Gamma^m_{sm} + g^{rs}g_{irk}\Gamma^p_{sp}\right]_{=0\,(ik)} \quad\quad\text{(III.D,5)}$$

$$+ g^{rs}\left\{ 2\Gamma^l_{si}\Gamma_{k,rl} - \Gamma^l_{il}\Gamma_{k,rs}\right\}$$

$$+ g^{rs}\left[-2g_{wks}\Gamma^w_{ri} + g^{pu}g_{iw}(g_{urk} - g_{rku})\Gamma^w_{sp} + \left(g^{xy}g_{ir}g_{ykx}\right)\Gamma^p_{sp}\right]$$

$$=$$

$$R_{ik} - \tfrac{1}{2} g_{ik}\left[R + 2G - \left(g^{ar}g^{bs}g_{lab}\Gamma^l_{rs} - g^{al}g^{bs}g_{lab}\Gamma^r_{sr}\right) + g^{rs}g^{al}(g_{sra} - g_{ars})\Gamma^m_{lm}\right]$$

$$2\Gamma^r_{is}\Gamma^s_{kr} - \Gamma^s_{is}\Gamma^r_{kr}$$

$$+ g^{rs}\left\{ 2\Gamma^l_{si}\Gamma_{k,rl} - \Gamma^l_{il}\Gamma_{k,rs}\right\}$$

$$+ g^{rs}\left[-2g_{wks}\Gamma^w_{ri} + g^{pu}g_{iw}(g_{urk} - g_{rku})\Gamma^w_{sp} + \left(g^{pu}g_{ir}g_{ukp}\right)\Gamma^m_{sm}\right] \quad\quad\text{(III.D,6)}$$

$$=$$

$$R_{ik} - \tfrac{1}{2} g_{ik}\left[R + 2G - g^{ar}g^{bs}g_{lab}\Gamma^l_{rs} + g^{al}g^{bs}(g_{lab} + g_{lab} - g_{bal})\Gamma^r_{sr}\right]$$

$$2\Gamma^r_{is}\Gamma^s_{kr} - \Gamma^s_{is}\Gamma^r_{kr} + g^{rs}\left\{ 2\Gamma^l_{si}\Gamma_{k,rl} - \Gamma^l_{il}\Gamma_{k,rs}\right\}$$

$$+ g^{rs}\left[-2g_{wks}\Gamma^w_{ri} + g^{pu}g_{iw}(g_{urk} - g_{rku})\Gamma^w_{sp} + \left(g^{pu}g_{ir}g_{ukp}\right)\Gamma^m_{sm}\right]$$

$$=$$

$$R_{ik} - \tfrac{1}{2}g_{ik}\Big[R + 2G - g^{ar}g^{bs}(g_{abl}+g_{lab}-g_{bla})\Gamma^{l}_{rs} + g^{al}g^{bs}(g_{abl}+g_{lab}-g_{bla})\Gamma^{r}_{sr}\Big]$$

$$-g^{rs}\Gamma^{m}_{im}\Gamma_{s,kr} + g^{rs}\Big\{\Gamma^{l}_{si}(g_{lkr}+g_{rlk}-g_{krl}) - \Gamma^{l}_{il}\Gamma_{k,rs}\Big\} \qquad \text{(III.D,7)}$$

$$+2g^{xs}\Gamma^{r}_{is}\Gamma_{x,kr} + g^{rs}\Big[-2g_{wks}\Gamma^{w}_{ri} + g^{pu}g_{iw}(g_{urk}-g_{rku})\Gamma^{w}_{sp} + \big(g^{pu}g_{ir}g_{ukp}\big)\Gamma^{m}_{sm}\Big]$$

in the 1. round brackets ($g_{abl}+g_{lab}-g_{bla}$) ← $g_{lab}$, the indices $a, b$ may have been exchanged [($rs$)-symmetry of factor $\Gamma^{l}_{rs}$]. In the 2. pair this does not apply [no ($ls$)-symmetry available], but with $l↔a$ because of factor $g^{al}$ there ½$g_{lab} = \Gamma_{b,la}$ is used now.

$$=$$

$$R_{ik} - \tfrac{1}{2}g_{ik}\Big[R + 2G - 2\big(g^{ar}\Gamma^{s}_{la}\Gamma^{l}_{rs} - g^{al}\Gamma^{s}_{la}\Gamma^{r}_{sr}\big)\Big]$$

$$-g^{rs}\Gamma^{m}_{im}\Gamma_{s,kr} + g^{rs}\Big\{\Gamma^{l}_{si}(g_{lkr}+g_{rlk}-g_{krl}) - \Gamma^{l}_{il}\Gamma_{k,rs}\Big\}$$

$$+g^{rs}\Big[2\Gamma^{w}_{ir}\Gamma_{s,kw} - 2g_{wks}\Gamma^{w}_{ri} + g^{pu}g_{iw}(g_{urk}-g_{rku})\Gamma^{w}_{sp} + \big(g^{pu}g_{ir}g_{ukp}\big)\Gamma^{m}_{sm}\Big]$$

$$=$$

$$R_{ik} - \tfrac{1}{2}g_{ik}[R]$$

$$-g^{rs}\big(\Gamma_{s,kr} + \Gamma_{k,rs}\big)\Gamma^{m}_{im}$$

$$+\Big[g^{rs}g^{pu}g_{iw}(g_{urk}-g_{rku})\Gamma^{w}_{sp} + g^{rs}\big(g^{pu}g_{ir}g_{ukp}\big)\Gamma^{m}_{sm}\Big]$$

$$=$$

$$R_{ik} - \tfrac{1}{2}g_{ik}R$$

$$-g^{rs}g_{rsk}\Gamma^{m}_{im}$$

$$+\Big[0 + g^{pu}g_{ukp}\Gamma^{m}_{im}\Big]$$

$$=$$

$$R_{ik} - \tfrac{1}{2}g_{ik}R \qquad \text{(III.D,8)}$$

As usual, among other conversions, here also e.g. $\Gamma_{k,li} = \tfrac{1}{2}g_{lik}$ is repeatedly used with regard to the ($ik$)-symmetry above.

$$\left\langle \frac{\partial \mathbf{G}^{(ik)}}{\partial g^{ik}} \right\rangle - \partial_{l}\left\langle \frac{\partial \mathbf{G}}{\partial g^{ik}_{l}} \right\rangle - g_{ia}g_{kb}\left\langle \frac{\partial \mathbf{G}_{(ab)}}{\partial g_{ab}} \right\rangle \;=\; \mathbf{R}_{ik} - \tfrac{1}{2}g_{ik}\mathbf{R} \;=\; \mathbf{E}_{ik} \qquad \text{(III.D,9)}$$

(q. e. d.)

## IV. Exposing the problematic form of Einstein's and Hilbert's expressions (27), (39)

Making use of the symbolic kind of 'tensor differentiation' $\langle(..)\rangle$ given in the text, then starting from the action integral (31) of the note again but taking the $\Gamma_{a,bc}$'s necessarily from (III,1) to have **G** as a function of $g_l^{ik}$ instead of $g_{lab}$, one will correspondingly find symbolic Euler-Lagrange expressions, now 'derived' from (31) implicitly by definition.

$$\int \delta \mathbf{G}\, d\Omega \equiv \int \left( \left\langle \frac{\partial \mathbf{G}}{\partial g^{ik}} \right\rangle \delta g^{ik} + \left\langle \frac{\partial \mathbf{G}}{\partial g_l^{ik}} \right\rangle \delta g_l^{ik} \right) d\Omega \ . \tag{IV,1}$$

After partial integration (with vanishing variations $\delta g^{ik}$ at the boundaries) this would yield

$$\left\langle \frac{\partial \mathbf{G}}{\partial g^{ik}} \right\rangle - \partial_l \left\langle \frac{\partial \mathbf{G}}{\partial g_l^{ik}} \right\rangle \tag{IV,2}$$

for $\mathbf{E}_i^{\ k}$, while Einstein's original expression (27) does not apply without changing the straightforward definition of partial derivatives correspondingly. By consequent further evaluation it follows according to (38) and using (13) appropriately

$$\mathbf{E}_{ik} = \frac{1}{X(ik)} \left[ \frac{\partial \mathbf{G}^{(ik)}}{\partial g^{ik}} - \partial_l \left( \frac{\partial \mathbf{G}}{\partial g_l^{ik}} \right) + \left( \frac{\partial g_{ab}}{\partial g^{ik}} \right) \frac{1}{X(ab)} \left( \frac{\partial \mathbf{G}_{(ab)}}{\partial g_{ab}} \right) \right] , \tag{IV,3}$$

where $\partial \mathbf{G}^{(ik)}/\partial g^{ik}$ and $\partial \mathbf{G}^{(ab)}/\partial g_{ab}$ obviously mean partial derivations with respect to the contravariant and covariant *tensor components* appearing in the same **G** now. Concerning the last summand of (38), however, the first of both $\langle(..)\rangle$-normalizations would be redundant if in

$$\frac{\partial g_{ab}}{\partial g^{ik}} = -\tfrac{1}{2} X (g_{ia} g_{kb} + g_{ka} g_{ib}) \tag{IV,4}$$

the factor $X$ would equal unity consistently [the $(ik)$-symmetrized form used here is only necessary to compare both sides of (IV,4) directly, otherwise already a factor like $(\partial \mathbf{G}/\partial g_{ab})$ in e.g. (IV,3) provides $(ik)$-symmetry (even before any summation is done)]. A direct calculation, however, shows that actually it is

$$\begin{aligned} X &= 1 \ \text{in case } (i=k) \\ X &= 2 \ \text{in case } (i \neq k) \end{aligned} \tag{IV,5}$$

what according to (11) here means

$$X = \left[ 1 + \text{sign}(|i-k|) \right] . \tag{IV,6}$$

Comparing (38) with the $(ik)$-symmetrized right hand side of equation (36), however,

$$\left\langle \frac{\partial \mathbf{G}^{(ik)}}{\partial g^{ik}} \right\rangle - \partial_l \left\langle \frac{\partial \mathbf{G}}{\partial g_l^{ik}} \right\rangle - \tfrac{1}{2}(g_{ia}g_{kb} + g_{ka}g_{ib}) \left\langle \frac{\partial \mathbf{G}_{(ab)}}{\partial g_{ab}} \right\rangle , \tag{IV,7}$$

which applies as proved in Section II above, one would have found (38) an equivalent result, only if (23) is adapted as the suitable definition suggested by the well-known relation (5), though in the proper meaning of conventional tensor analysis a corresponding relation '$X_{ikmn} = \partial g_{mn}/\partial g^{ik}$' of unclear character does neither exist nor, to my knowledge, has been used by Hilbert or Einstein. It seems anything but a purely coincidental aspect, that relation (IV,4) – in contrast to (5) – may not have been used in the literature so far. Only by definition of (23), relation (36) can be formally reduced to Einstein's modified expression (IV,2).

Concerning a direct component calculation of Hilbert's formula (39), who in [2] has effectively stated the numerical factors $X$ and $Y$ for partial derivatives with respect to $g^{ik}$, $g_l^{ik}$, or $g_{lm}^{ik}$ — in general according to (11), (12) – relation (48) may be explicitly written in the form

$$\mathbf{E}_{ik} = \frac{1}{X(ik)}\left[\frac{\partial \mathbf{R}^{(ik)}}{\partial g^{ik}} - \partial_l\left(\frac{\partial \mathbf{R}}{\partial g_l^{ik}}\right)\right] + \frac{1}{Y(lm,ik)}\partial_{lm}\left(\frac{\partial \mathbf{R}}{\partial g_{lm}^{ik}}\right) + \frac{1}{X(ik)}\left(\frac{\partial g_{ab}}{\partial g^{ik}}\right)\frac{1}{X(ab)}\left(\frac{\partial \mathbf{R}_{(ab)}}{\partial g_{ab}}\right), \quad \text{(IV,8)}$$

where mixed 'Hilbert factors' $1/X$ and $1/Y$ as well as products of various factors are included, which should have made such a calculation nearly impracticable if not only in the sense of the symbolic 'tensor differentiations' according to (48) in this case, thus actually representing an application of the principle of least action instead of the conventional Euler-Lagrange formalism. Now, nevertheless, the rather circumstantial derivations (38), (IV,3), or (48), (IV,8) of Einstein's tensor density (1) using the merely symbolic Euler-Lagrange formalism seem to yield the first quasi-direct calculations corresponding to the unnecessarily problematic expression (27), (39) after all.